\documentclass[preprint, authoryear, 5p, number]{elsarticle}
\usepackage{hyperref}
\usepackage{units}
\usepackage{ amssymb }
\usepackage[english]{babel}
\usepackage{color}
\usepackage{array}
\usepackage{booktabs}
\usepackage{graphicx}
\usepackage{array}

\newcolumntype{M}[1]{>{\centering\arraybackslash}m{#1}}
\newcolumntype{N}{@{}m{0pt}@{}}

\journal{Astronomy and Computing}
 
\begin{document}
 
\begin{frontmatter}
 
\title{The role of 3-D interactive visualization in blind surveys of H\,{\large I} in galaxies}

 \author[a]{D.~Punzo\corref{cor1}}
 \ead{D.Punzo@astro.rug.nl}
 
 \author[a]{J.M.~van der Hulst}
 \author[b]{J.B.T.M.~Roerdink}
 \author[a,c]{T.A. Oosterloo}
 \author[a]{M. Ramatsoku}
 \author[a]{M.A.W. Verheijen}

 \cortext[cor1]{Corresponding author}

 \address[a]{Kapteyn Astronomical Institute, University of Groningen, P.O. Box 800, 9700 AV Groningen, The Netherlands}
  
 \address[b]{Johann Bernoulli Institute for Mathematics and Computer Science, University of Groningen, Nijenborgh 9, 9747AG Groningen, The Netherlands}
 
 \address[c]{ASTRON, the Netherlands Institute for Radio Astronomy, Postbus 2, 7990 AA, Dwingeloo, The Netherlands}

\begin{abstract}

Upcoming H\,{\small I} surveys will deliver large datasets, and automated processing using
the full 3-D information (two positional dimensions and one spectral dimension) to find and 
characterize H\,{\small I} objects is imperative. In this context, visualization is an essential 
tool for enabling qualitative and quantitative human control on an automated source finding and 
analysis pipeline. We discuss how Visual Analytics, the combination of automated data processing 
and human reasoning, creativity and intuition, supported by interactive visualization, enables 
flexible and fast interaction with the 3-D data, helping the astronomer to deal with the analysis 
of complex sources. 3-D visualization, coupled to modeling, provides additional capabilities 
helping the discovery and analysis of subtle structures in the 3-D domain. The requirements for a 
fully interactive visualization tool are: coupled 1-D/2-D/3-D visualization, quantitative and comparative 
capabilities, combined with supervised semi-automated analysis. Moreover, the source code must have 
the following characteristics for enabling collaborative work: open, modular, well documented, and 
well maintained. We review four state of-the-art, 3-D visualization packages assessing their 
capabilities and feasibility for use in the case of 3-D astronomical data.

\end{abstract}
 
 
\end{frontmatter}

\section{Introduction}

The Square Kilometre Array (SKA) and its precursors are opening up new opportunities for radio 
astronomy in terms of data collection and sensitivity. Two types of blind surveys are 
planned with SKA-pathfinders:

\begin{enumerate}
\item shallow (very large sky coverage): WALLABY with ASKAP \citep{askap, duffy}, 
shallow and medium-deep APERTIF surveys with the WSRT \citep{Apertif3}.

\item deep (high sensitivity, small solid angle): CHI-LES with the J-VLA \citep{vla, chiles}; 
LADUMA with MeerKAT \citep{meerkat,laduma} and DIN-GO with ASKAP \citep{askap, duffy}.
\end{enumerate}

The first type of H\,{\small I} surveys will detect $\sim 10^3$ sources weekly, of which $0.2 \%$ 
will consist of well resolved sources, $6.5 \%$ will have a limited number of resolution elements, and 
$93 \%$ will at best be marginally resolved \citep{duffy}. This predicted weekly data rate is 
high, and fully automated pipelines will be required for processing the data (see section 
\ref{Large}). The first and second category of sources will contain a wealth of morphological and 
kinematic information. However, in cases with complex kinematics it will be difficult to extract all 
information in a controlled and quantitative way \citep{Sancisi, Boomsma}. Therefore, manual analysis 
of a subset of the resolved sources will still be required. In fact, manual processing will be 
very useful for obtaining a deeper insight in particular features of the data (e.g., tails and 
extra-planar-gas; see section \ref{3-D}). It will also enhance possible improvements to the 
automated pipelines. For example, it can play a major role in the development and training of machine 
learning algorithms for the automated analysis data, in particular in the era of the full SKA data 
(see section \ref{ana}).

The SKA pathfinders will provide a wealth of data, but the expected exponential growth 
of the data has created several data challenges. We will present a preview of the data that 
APERTIF will deliver to the community in the near future and discuss the importance of 
visualization for the analysis of radio data in the upcoming surveys era. Our 
discussion will be based on existing mosaics acquired with the Westerbork Synthesis Radio Telescope (WSRT), 
which are representative for the daily image data rate provided by future blind 
H\,{\small I} surveys.

\subsection{WSRT and the APERTIF data}

The WSRT consists of a linear array of 14 antennas with a diameter of 25 meters arranged on a 
2.7 km East-West line located in the north of the Netherlands. The APERTIF phased array feed 
system is an upgrade to the WSRT which will increase the field of view by a factor of 30 
\citep{apertif2, apertif1}, which allows a full inventory of the northern radio sky complemented 
by a wealth of optical, near-IR data, and other radio observatories such as the Low-Frequency 
Array (LOFAR). Part of the APERTIF surveys will be a medium deep blind survey of a few 
hundred square degrees with a 3$\sigma$ column density depth of $2-5 \times 10^{19}$ cm$^{-2}$.

A full 12 hour integration will provide $\sim 2.4$ TB of \textit{complex visibilities} 
sampling a $3^{\circ}\times 3^{\circ}$ region of the sky and the following data reduction will 
generate three dimensional data sets of the H\,{\small I} line emission, with axes right 
ascension (RA or $\alpha$), declination (DEC or $\delta$), and frequency ($\lambda$) or recession 
velocity ($v$). The typical size of a data cube will be 2048 $\times$ 2048 pixels for the  spatial 
coordinates (each pixel covers 5 arcsec) and 16384 spectral channels, which correspond to 16384 pixels 
in the third dimension covering a bandwidth of 300 MHz ($\sim 60,000$ km/s). The disk storage needed 
for each data cube is about 0.25 TB, assuming a single Stokes component, I, and 
32 bits per pixel format. The final product after observing the northern sky will be of 
the order of 5 PB of data cubes. 

Examining these numbers it is clear that the storage, data 
reduction, visualization and analysis to obtain scientific results requires the development of 
new tools and algorithms which must exploit new solutions and ideas to deal with this large 
volume of data. The Tera-scale volume of these datasets produces, in fact, both technical issues 
(e.g., dimension of the data much larger than the available random access memory (RAM) on a 
normal workstation) and visualization challenges (i.e., the presence in each dataset of a large 
number of small sources with limited signal-to-noise ratio (SNR)).

\subsection{Data visualization}

Traditionally visualization in radio astronomy has been used for:
\begin{enumerate}[i)]
\item finding artefacts due to an imperfect reduction of the data;\label{i}                          
\item finding sources and qualitatively inspect them;\label{ii} 
\item performing quantitative and comparative analysis of the sources.\label{iii} 
\end{enumerate}

In this paper we will focus mainly on the connection between interactive visualization 
and the automated source finder and analysis pipeline (\ref{ii}); and the importance of 
interactive, quantitative and comparative visualization (\ref{iii}). We will not 
discuss visualization of artifacts (\ref{i}) resulting from imperfections in the data. Artifacts 
can arise from several effects: Radio Frequency Interference (RFI), errors in the bandpass 
calibration, or errors in the continuum subtraction. Volume rendering can help localizing such 
artifacts, but in that case visualization is envisaged to play the role of assisting quality control of 
the products of an automated calibration pipeline. This will be the subject of a separate study 
as it may require different tools.

In section \ref{astro} we give an overview of the past and current visualization
packages and algorithms, with a focus on radio astronomy. We highlight the 3-D nature of the 
H\,{\small I} data in section \ref{Large}. The definition of the requirements for a fully interactive
visualization tool is given in section \ref{prere}. Finally, in section \ref{review_sof}, we 
review state of-the-art visualization packages with 3-D capabilities. Our aim is to define the 
basis for the development of a 3-D interactive visualization tool.

\section{Scientific visualization}\label{astro}

Scientific visualization is the process of turning numerical scientific data into a visual 
representation that can be inspected by eye. The concept of scientific visualization born  
in the 80's \citep{McCormick,Frenkel,DeFanti}. Its role was not relegated to only presentation \citep{roerdink}.
The interactive processing of the data, the imaging and analysis, including qualitative, 
quantitative and comparative stages, is crucial for archiving a deep and complete knowledge. 

In this section we provide background information about past visualization 
developments in astronomy, scientific visualization theory, visualization hardware 
and the software terminology used in this paper.

\subsection{Visualization in astronomy}

One of the first systematic radio astronomy visualization trials was undertaken by 
\cite{Allen1} \citep[see also][]{Allen2, Allen3, Allen4}. They investigated techniques for 
displaying single-image data sets, including contour display, ruled surface display, grey scale 
display, and pseudo-color display. They also discussed techniques for the display of multiple 
image data sets, including false-color display and cinematographic display.

At the beginning of the 90's,  \cite{Mickus1}, \cite{Domik1}, \cite{Mickus2}, \cite{Domik2}, and
\cite{Brugel} developed a visualization tool named the Scientific Toolkit for Astrophysical Research
($\tt{STAR}$). $\tt{STAR}$ was a prototype resulting from the development of an user interface and 
the implementation of visualization techniques suited to the needs of astronomers at that time. 
These included display of one- and two-dimensional datasets, perspective projection, 
pseudo-coloring, interactive color coding techniques, volumetric data displays, and data slicing.

Recently, both \citet{Hassan3} and \citet{Koribalski} pointed out the lack of a tool 
that can deal with large astronomical data cubes. In fact, the current astronomy 
software packages are characterized by a window interface for 2-D visualization of slices 
through the 3-D data cube; in some cases limited 3-D rendering is also present. Moreover, they 
can exploit only the resources of a personal computer which imposes strong limitations on the 
available RAM and processing power. Stand-alone visualization tool examples are $\tt{KARMA}$ 
\citep{Karma}, $\tt{SAOImage \, DS9}$  \citep{DS9}, $\tt{VisIVO}$ \citep{visivo, visivo1} and 
$\tt{S2PLOT}$ \citep{S2PLOT}. Other viewers exist but are embedded in reduction and analysis 
packages: $\tt{GIPSY}$ \citep{Gipsy, Gipsy1}, $\tt{CASA}$ \citep{CASA} and $\tt{AIPS}$ 
\citep{AIPS}. A recent development is the use of the open source software $\tt{Blender}$ 
for visualization of astronomical data \citep{Kent, Taylor}, but this application is more 
suitable for data presentation rather than interactive data analysis. 

From the inventory of the current state of-the-art we conclude that the expected 
exponential growth of radio astronomy data both in resolution and field of view has 
created a necessity for new visualization tools. In the meantime much development 
has taken place in computer science and medical visualization. We review relevant software  
from these areas in sections \ref{Large} and \ref{review_sof}.

\subsection{3-D visualization}

First investigations of the suitability of 3-D visualization 
for radio-astronomical viewers date back to the beginning of the 90's \citep{Norris}.
Already at that time it was clear that a 3-D approach can provide a better understanding
of the 3-D domain of the radio data. The type of data slicing commonly used 
(i.e., channel movies), forces the researcher to remember what was seen in other channels
and requires a mental reconstruction of the data structure. The major advantage of a 3-D
technique is an easier visual identification all structure, including faint features 
extending over multiple channels. A crucial point made by Norris is that presenting 
the results qualitatively is fine for data inspection, but that interactive and 
quantitative hypothesis testing requires quantitative visualization.

In the last twenty years hardly any new 3-D visualization tools were developed
for examining 3-D radio astronomical data. In the middle of the 90's, \citet{Oosterloo} 
investigated porting direct volume rendering techniques to radio astronomy 
visualization. He analyzed the features and the issues related to a ray casting algorithm (a 
massively parallel image-order method, see \citet{Roth}), pointing out, in general, the 
advantages and drawbacks of the 3-D visualization. He could, however, not develop a run-time 3-D 
interactive software package due to the lack of available computational resources.

\subsection{Volume rendering}

3-D visualization is the process of creating a 2-D projection on the screen of the 3-D objects 
under study. This process is called volume rendering. The rendering methods are divided in two 
principal families: indirect volume rendering (or surface rendering) and direct volume rendering.
The first approach fits geometric primitives through the data and then it renders the image. It 
requires a pre-processing step on the dataset, then after the pre-processing a quick 
rendering is possible. Fitting geometric primitives, however, may introduce noise errors due to  
rendering artifacts. Moreover, not all datasets can be easily approximated with geometric 
primitives and the H\,{\small I} sources fall into this class because they do not have well defined 
boundaries. Furthermore, in a H\,{\small I} data cube the signal-to-noise is usually low. For 
example, the galaxies in the WHISP survey \citep{Whisp}, have average signal-to-noise of $\sim$ 10 in the 
inner parts and $\sim$ 1 in the outer parts. This makes indirect volume rendering inefficient. Direct 
volume rendering methods directly render the data defined on a 3-D grid (each element of the grid is 
called a \textit{voxel}), therefore it requires more computations to process an image. Several 
direct rendering solutions exist and they are classified as: 1) Object order methods, requiring 
an iteration over the voxels which are projected on the image plane; 2) Image order methods,
which instead iterate over the pixels of the final rendered image 
and have the algorithm calculate how 
each voxel influences the color of a single pixel. 3) Hybrid methods, a combination of 
the first two. It must be noted that during the process of direct volume rendering the depth 
information can be mixed depending on the projection method used (i.e., \textit{maximum, 
minimum,} and \textit{accumulate}). By rotating or the use of 3-D hardware the human user is 
able to mentally connect the various frames and to register the proper 3-D structures. For a detailed 
review of the state of-the-art and for more information we refer to the Visualization Handbook 
\citep{Hansen} and the VTK book \citep[4th edition]{Schroeder}.

\subsection{Out-of-core and in-core solutions}

The rendering software can exploit an out-of-core or an in-core solution. Out-of-core solutions 
are optimized algorithms designed to handle datasets larger than the main system memory by 
utilizing secondary, but much slower, storage devices (e.g., hard disk) as an auxiliary memory layer. 
These algorithms are optimized to efficiently fetch or pre-fetch data from such secondary 
storage devices to achieve real-time performance. They usually utilize a multi-resolu\-tion data 
representation to facilitate such a fast fetch mechanism and to support different available 
output resolutions based on the limitations in terms of the processing time and the available 
computational resources \citep{Rusink,Crassin}. 

The in-core solution can achieve very fast memory transfer because it does not need to access 
the data stored on a hard disk continuously. In fact, it assumes that the data are stored in 
the main system memory, ready for processing. Of course, in this case the main limitation is 
the size of the available RAM.

\subsection{3-D hardware}\label{vireal}

The use of 2-D input and output hardware limits the possible interaction with a 3-D 
representation. 3-D input devices (such as 3-D mouse or pointer) can naturally solve this 
problem. Moreover, coupling them to a 3-D output device such as a 3-D monitor, a CAVE virtual 
environment, etc., can remove the difficulty of positioning a 3-D cursor in a 3-D space. In fact,
in this case, the user can see the real 3-D movement, instead of the projection on a 2-D screen.
However, virtual reality has never been widely used in the researchers' daily work due to the 
dependence on very expensive hardware not available on the common computer market.

Recently, two new very promising devices, the Leap Motion (an input device that tracks the 
hands in 3-D \footnote{\url{https://www.leapmotion.com/}}) and the Oculus Rift (a 3-D output 
device with a full immersion virtual reality experience \footnote{\url{http://www.oculus.com/}}), 
appeared that can change this situation, because they are aimed for the gaming market, 
and therefore will be rather cheap.

This hardware could enhance new interaction perspectives with volumetric data using a desktop 
solution. We will however exclude them from our visualization discussion because the success 
and therefore the maintainability of a visualization solution based on them, which depends on the 
gaming market, is still uncertain. Moreover, from the point of view of interface design, the use 
of this new hardware creates the need to develop new interface concepts. The equivalent expertise 
that exists for classical interfaces such as mouse and keyboard is, however, still missing. This 
does not exclude that in the coming years virtual reality may become very popular and stimulate 
many developers to experiment with the Leap Motion and the Oculus Rift or future 3-D 
hardware.

\subsection{Visual Analytics}\label{ana}

In the SKA era manual inspection and analysis of even only
  a subset of data will be extremely hard to achieve. Machine learning 
  will be needed for
  classification of the different components of a galaxy 
  \citep{calleja, banerji}. However, the reliability of the
  analysis by machine learning heavily depends on the input for the
  training session \citep{kuminski}. Discovering interesting
  relations, structures, and patterns in very large and
  high-dimensional data spaces needs the combination of automated data
  processing with human reasoning, creativity and intuition, supported
  by interactive visualization. Human assessment remains essential for
  understanding the behavior of automatic algorithms and for visual
  quality control. As the available data grow, effective and
  efficient techniques are essential to increase our insight in the
  underlying structures and processes.

Combining interactive visualization with analytic techniques
  (machine learning, statistics, data mining) has grown into a field
  of its own: \emph{Visual Analytics}~\citep{thomas05:_illum_path,
    keim10:_master}.  It aims to fully integrate human expertise in
  the human-machine dialogue to steer the sense-making process. Visual
  analytics supports collaborative exploration and decision making by
  combining fast access to large distributed databases, data
  integration, powerful computing infrastructures, and interactive
  visualization facilities (\emph{e.g.}, large touch displays).
  Astronomy is an exciting and extremely demanding testfield for new
  visual analytics techniques. Data availability, storage and
  distribution are well covered.  Expert knowledge is available to
  validate algorithmic approaches.  Data-set dimensionality (dimension
  $d=10\ldots 100$) and sizes ($>10^9$ elements) make scalability
  extremely difficult to achieve.  Extracting meaningful relations
  across the \emph{entire} set of data dimensions is inherently hard
  for data of high
  dimensionality ~\cite{bertini11:_qualit_metric_high_dimen_data_visual}. 
  Integrating data sources, data-reduction algorithms, and expert knowledge to
  effectively and efficiently answer domain-specific questions is an
  open challenge. Visual analytics advocates a mixed approach:
  automatically search datasets for potentially meaningful patterns,
  and interactively steer data reduction and visualization. 

\section{Visualization of H\,{\small I} data sets}\label{Large}

The domain of future radio surveys, such as those plan-ned with APERTIF, will fall in the Big 
Data domain for two reasons:

\begin{enumerate}[i)]
\item a data cube will have dimensions of $2048\times2048\times16384$ $\sim 68.7 \times 
10^9$ (0.25 TB). The data  rate is $\sim 10$ cubes/week;
\item each data cube will contain $\sim 100$ sources, i.e. galaxies, of relatively small typical 
size ($\sim 10^{5}$  voxels) in the observed data volume of $ \sim 10^{11}$  voxels. 
\end{enumerate}

A very important step is to condense this vast amount of data collected by the surveys into 
a much smaller catalog of interesting regions, the sources, and their properties. This is 
done by examining the data itself. If done manually, the astronomers have to explore the whole 
data set using visualization software in order to identify the sources.
 
\subsection{Visualization and source finding}

For illustrative purposes, we consider a mosaicked data cube that serves
as a pilot training set for future, single Apertif pointings \citep[in prep.]{Mpati}.  
The mosaic is built from 35 individual WSRT
pointings in a hexagonal grid, directed towards a region in the sky
where a filament of the Perseus-Pisces Supercluster (PPScl) crosses the
plane of the Milky Way.  The data cube covers a sky area of
$2.4^{\circ} \times 2.4^{\circ}$ centered at $\alpha = 72^{\circ}$ and $\delta = 45^{\circ}$.
The redshift range is $c\,z$ = 2000 - 17000 km/s. The resulting data cube has dimension 2300
$\times$ 2300 pixels for the spatial coordinates and 1717 pixels in the velocity dimension. 
This is $\sim 10$ times smaller in the velocity (frequency) dimension than a single APERTIF
pointing, but the spatial resolution, velocity resolution and sensitivity are comparable. The
number of objects is also comparable as Perseus-Pisces is an over-dense region. The $\sim 200$
sources comprise $\lesssim 1\%$ of the data volume. The minimum column density detected
is $\sim 6.4 \times 10^{19}$ {cm$^{-2}$} at the $3\sigma$ level over a 
velocity range of 16.5 km/s.

The three-dimensional visual representation of the mosaic in Fig.\ref{fig0}
immediately highlights the sources' 
shape and position in the data cube. Moreover, interactivity such as rotation, zooming and 
panning, and editable color transfer functions greatly support manual identification of the 
sources in the data cube.

An interactive in-core ray casting algorithm running on a cluster of \textit{Graphic Processing 
Units} (GPUs) has been proposed by \citet{Hassan2} for the visualization of Tera-scale 
radio astronomy data cubes. In general, many large visualization software tools are in 
development in the context of computer science and medical imaging. Some notable 
examples follow:

\begin{enumerate}[i)]
\item in-core solutions exploiting parallel computing on a cluster:  \cite{ParaView} (i.e., 
$\tt{ParaView}$) and \cite{Vo};
\item out-of-core solutions: \cite{Crassin} (i.e., $\tt{Giga}$-\\ $\tt{Voxels}$) 
and \cite{Hadwiger}.
\end{enumerate}

In the case of visualization of H\,{\small I} in galaxies, it is, however, unlikely that visualization of 
the full data cube will be used for finding sources for the following reasons:

\begin{figure*}
\centering

\includegraphics[width=1.0\textwidth]{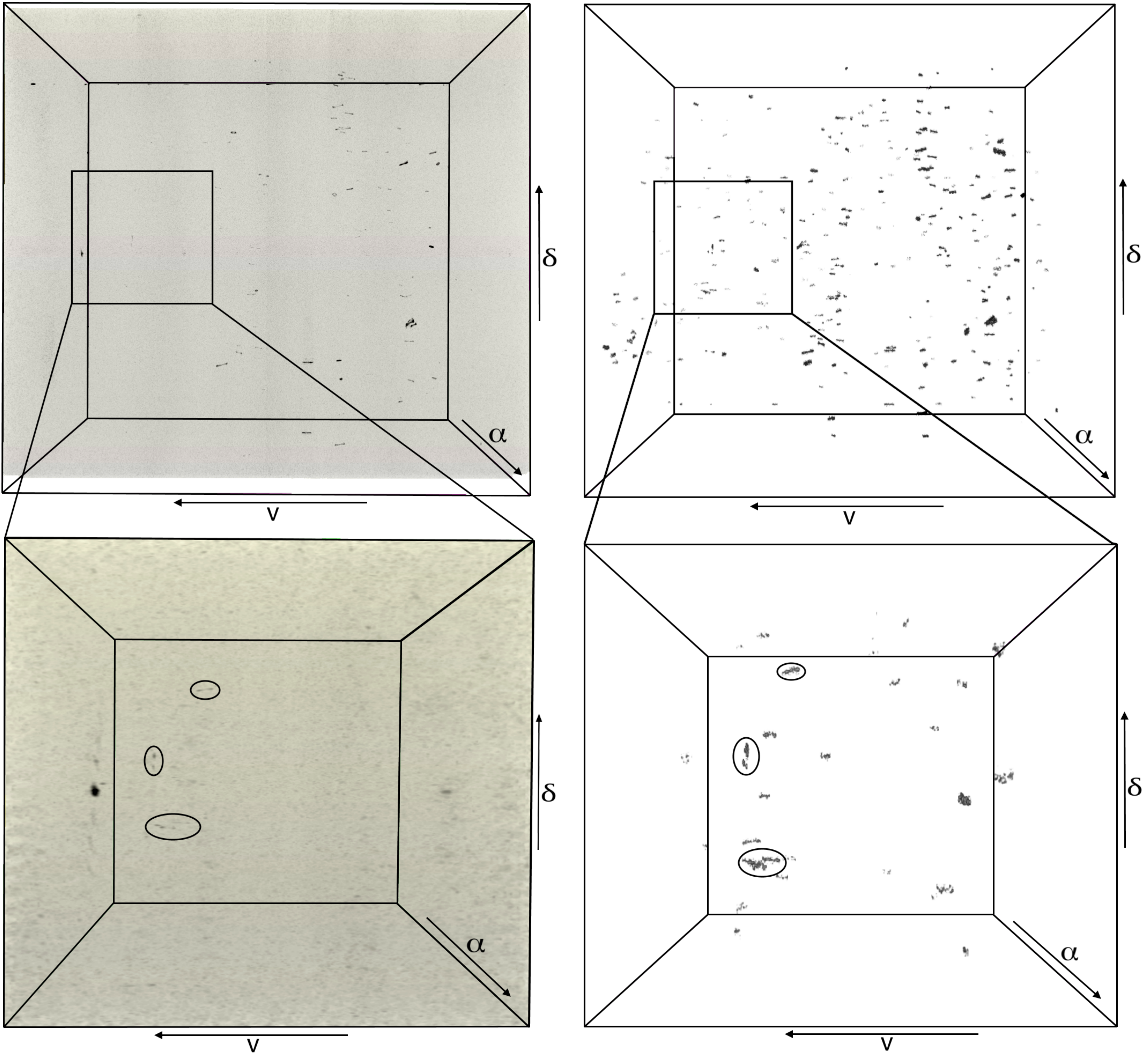}
\caption{Two representations of the H\,{\small I} in galaxies in a filament of the Perseus-Pisces 
Supercluster (PPScl) are shown.  In the top-left panel, the rendering of the full 
data cube with a maximum intensity projection method is illustrated. In the top-right 
panel, the data cube after a semi-automated procedure, performing, with $\tt{GIPSY}$ routines, 
the smooth and clip procedure as implemented in \cite{sofia}, is shown. In both 
cases it is possible to see a large number of sources, but many of them are hardly visible in the 
top-left panel. In the bottom panels, two zooms are provided. Smoothing has been applied at the 
bottom-left sub-cube revealing some of the sources (circled) easily visible in the bottom-right 
panel.} 
\label{fig0}
\end{figure*}

\begin{enumerate}[1)]
\item the size of the H\,{\small I} blind survey data volume and the number of 
sources, as illustrated in Fig.\ref{fig0}, prohibit a manual approach even 
when using very powerful interactive visualization tools;

\item radio data are intrinsically noisy, and most sources are faint and often extended.
Spatial and/or spectral smoothing increase the signal-to-noise ratio depending on the source 
structure. In fact, smoothing is applied on multiple spatial and spectral scales to ensure that sources 
of different size are extracted at their maximum, integrated signal-to-noise ratio. 
In Fig.\ref{fig0} two visual representations of the PPScl data, respectively before and after 
the source finder step, are shown. In both cases it is possible to see a large number of 
sources, but many of them are hardly visible in the original data cube because they 
drown in the noise. Manual operations such as zooming, changing the color function, and smoothing 
help the observer in identifying the sources visually. This will, however, take a prohibitive amount of 
time if done manually and will be impossible to perform if such data cubes are delivered at a rate of 1-2 
per day;

\item interactive rendering of $ \sim 10^{11}$  voxels using an in-core solution, such as 
\cite{Hassan2} demonstrated, requires considerable resources for hardware and maintenance, not affordable by 
typical research groups or major observatories. An out-of-core solution can reduce the 
financial demands on hardware. However, the development itself of such a solution requires a 
huge programming effort due to many challenges related to the I/O bandwidth limits. We 
refer to \cite{Crassin} and \cite{Hadwiger} for a detailed description of state of-the-art 
out-of-core visualization algorithms, including CPU-GPU memory transfer solutions. Note, 
however, that none of the rendering pipelines cited here are publicly available yet.
\end{enumerate}

Automated pipelines have been developed to extract the source information from the data 
collected \citep{Whiting,sofia}. Their goal is to find all reliably
detectable extragalactic H\,{\small I} objects in the observed data volume, 
and to determine the properties of these objects, that is:

\begin{enumerate}[a)]
\item the galaxies, i.e., the regularly rotating gas disks;\label{a}

\item additional H\,{\small I} structures such as extra-planar-gas and tails. These are crucial
for understanding the detailed balance between gas accretion and gas depletion processes, as well as their
dependence on the environment, and for obtaining the full picture of galaxy evolution. For example,
extra-planar-gas data can be used to quantitatively constrain the gas accretion and depletion
processes (see section \ref{analysis}). Another example is the presence of tails in the
data. Tails can be produced by tidal interactions between galaxies (Fig. \ref{fig02}) or by ram
pressure stripping \citep{Gorkom}, and are strong indications for these processes. Deciding which
process is important requires detailed inspection and modeling of the features discovered in
the data. We refer to \cite{Sancisi} for a full review of the state of-the-art of
H\,{\small I} observations and their interpretation. These features are located in the vicinity of the
galaxies and have low column densities and low signal-to-noise;\label{b}

\item the faint H\,{\small I} in the cosmic web such as H\,{\small I} filaments between galaxies. This 
emission is expected to have very low column density and very low signal-to-noise in a 
single resolution element, so will be difficult to detect. It is probably extended, following the 
large-scale structure, so the signal-to-noise could be increased by smoothing. This is, however, unlikely to be 
sufficient for detection (see below).\label{c}

\end{enumerate}

For inspecting (\ref{a}) and (\ref{b}), visualization techniques can be used in the following
approach: high-dimensional visualization (e.g., 3-D scatter plots) of the parameters provided by
the pipeline and stored in catalogs (such as position, flux, flux error, degree of
asymmetry, velocity width, integrated profile asymmetry, etc.) gives an overview of the data and
their 3-D domain (see section \ref{highvisu}). Then, manual inspection will be
performed for only a subset of sources, which can be delivered to
a visualization analysis package with full rendering capabilities for further analysis (see
section \ref{3-D}). In the case of (\ref{c}) we should point out that future 
observations with the SKA precursors, such as APERTIF and ASKAP, will not achieve the 
sensitivity to detect the cosmic web. The neutral 
fraction of cosmic web filaments is expected to be very low, leading to H\,{\small I} column densities 
$\lesssim 10^{18}$ cm$^{-2}$, \cite{Braun, ribaudo}. We therefore do not focus on such low level and
extended emissions.

\subsection{Automated pipelines and human intervention.}  
Automated pipelines will be responsible for finding the sources, measuring parameters that give 
an indication of the properties of a source and creating catalogs. Source finders are 
designed to automatically detect all the sources in the field. In order to do that, source 
finders must employ an efficient mechanism to discriminate between such interesting regions and 
the noise. The peak flux, total flux, and number of voxels are parameters that can be
used to determine the completeness and reliability of detected sources when examining both 
positive and negative detections \citep{serra}. Due to the 
complex 3-D nature of the sources \citep{Sancisi} and the noisy character of the data, it is, 
however, not trivial to construct a fully automated and reliable pipeline. A review of the 
current state-of-the-art is given by \citet{Popping}, who describe the issues connected with 
the noisy nature of the data, and the various methods and their efficiency. 
In addition, automated source characterization and measurement of source parameters 
are required for producing catalogs with science-ready products.
Human inspection will be necessary for quality control of the 
results from the pipelines and in particular for investigating complex cases. 
The human mind, in fact, is a very powerful diagnostic instrument which can 
naturally recognize (source) structures in the data. For 
example, in a significant number of cases, it will be very difficult to automatically 
retrieve information about particular features such as tidal tails or 
stripped H\,{\small I}. 
APERTIF most likely will deliver 2 or 3 of these cases every day (estimate based 
on the data shown in Fig.\ref{fig0}). The analysis of these will still be done manually and 
visualization will still play a major role. In fact, automated algorithms are built on 
the knowledge acquired during the manual approach (see section \ref{ana} for the role of 
interactive visualization and machine learning in visual analytics). Moreover, coupling 
visualization tools with semi-automated data analysis techniques is necessary in order to 
improve the inspection itself.  

The subcubes containing the sources detected by source finders will be relatively small with 
maximum sizes of $512\times512\times256 \sim 0.067 \times 10^9$ voxels, reducing the local 
storage, I/O bandwidth, and computational demand for visualization (easily achievable on a 
modern computer).

\subsection{Visualization and source analysis}\label{3-D}
\begin{figure}
\centering
\includegraphics[width=0.36\textwidth]{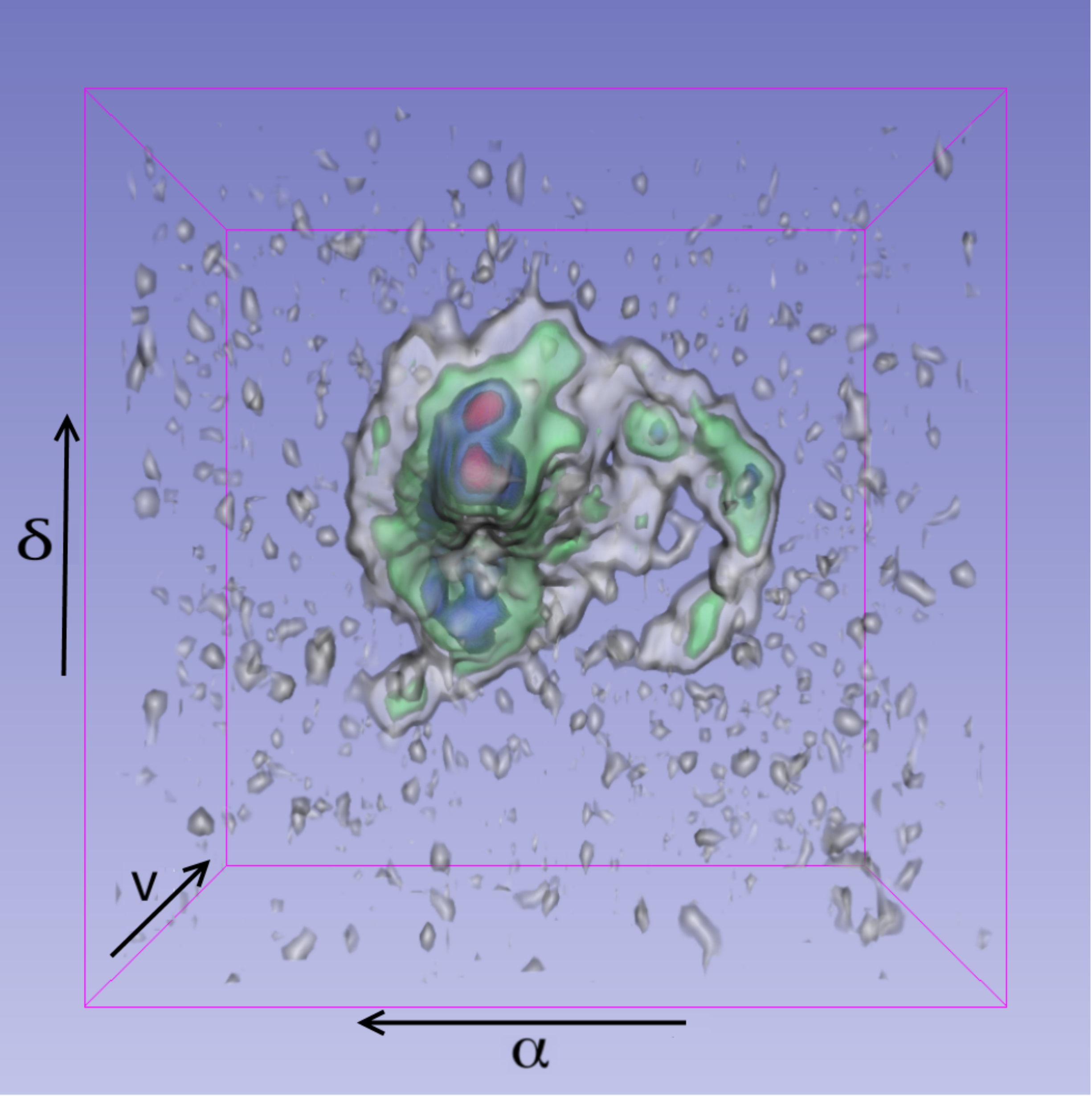}
\includegraphics[width=0.36\textwidth]{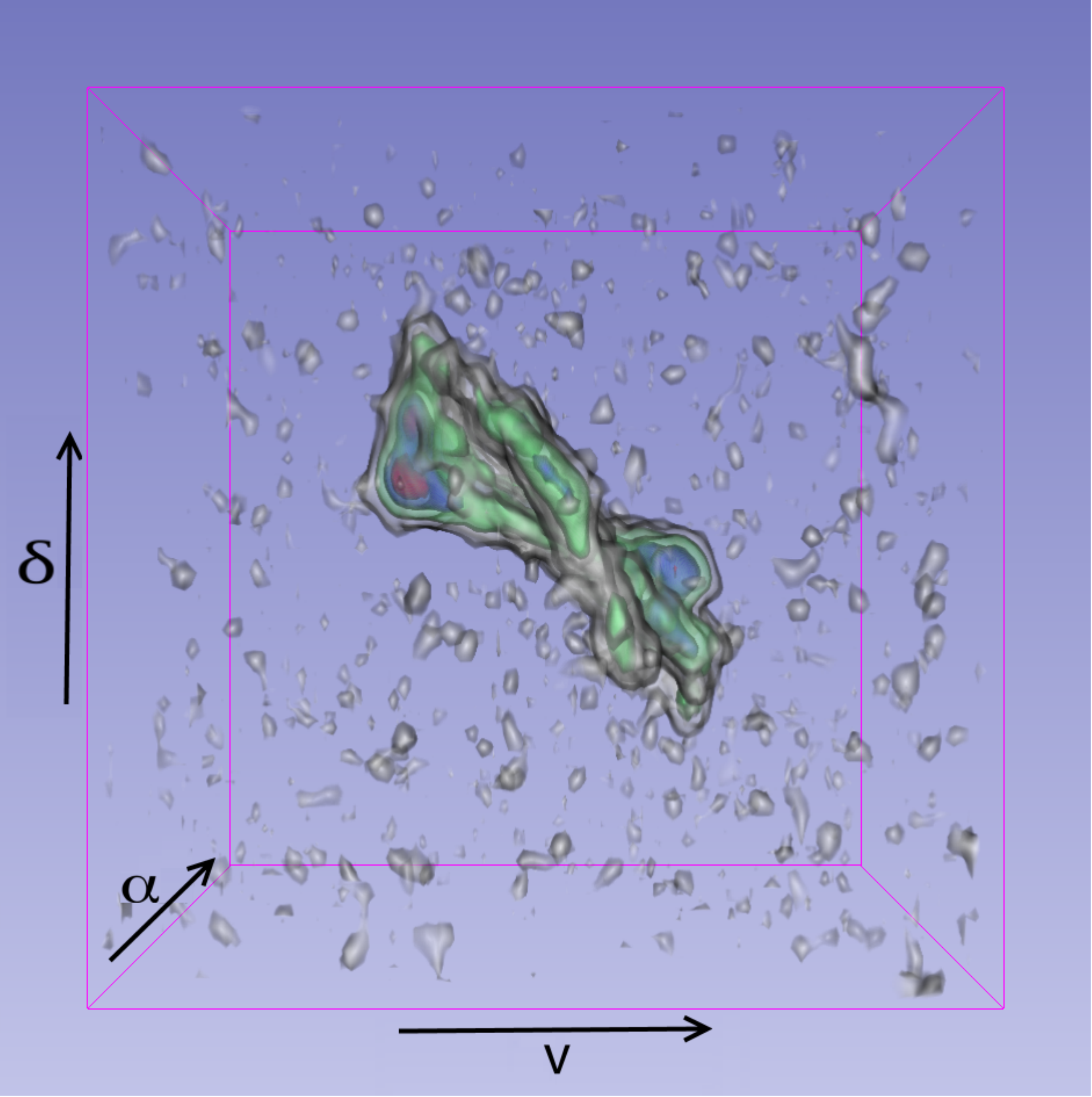}
\includegraphics[width=0.36\textwidth]{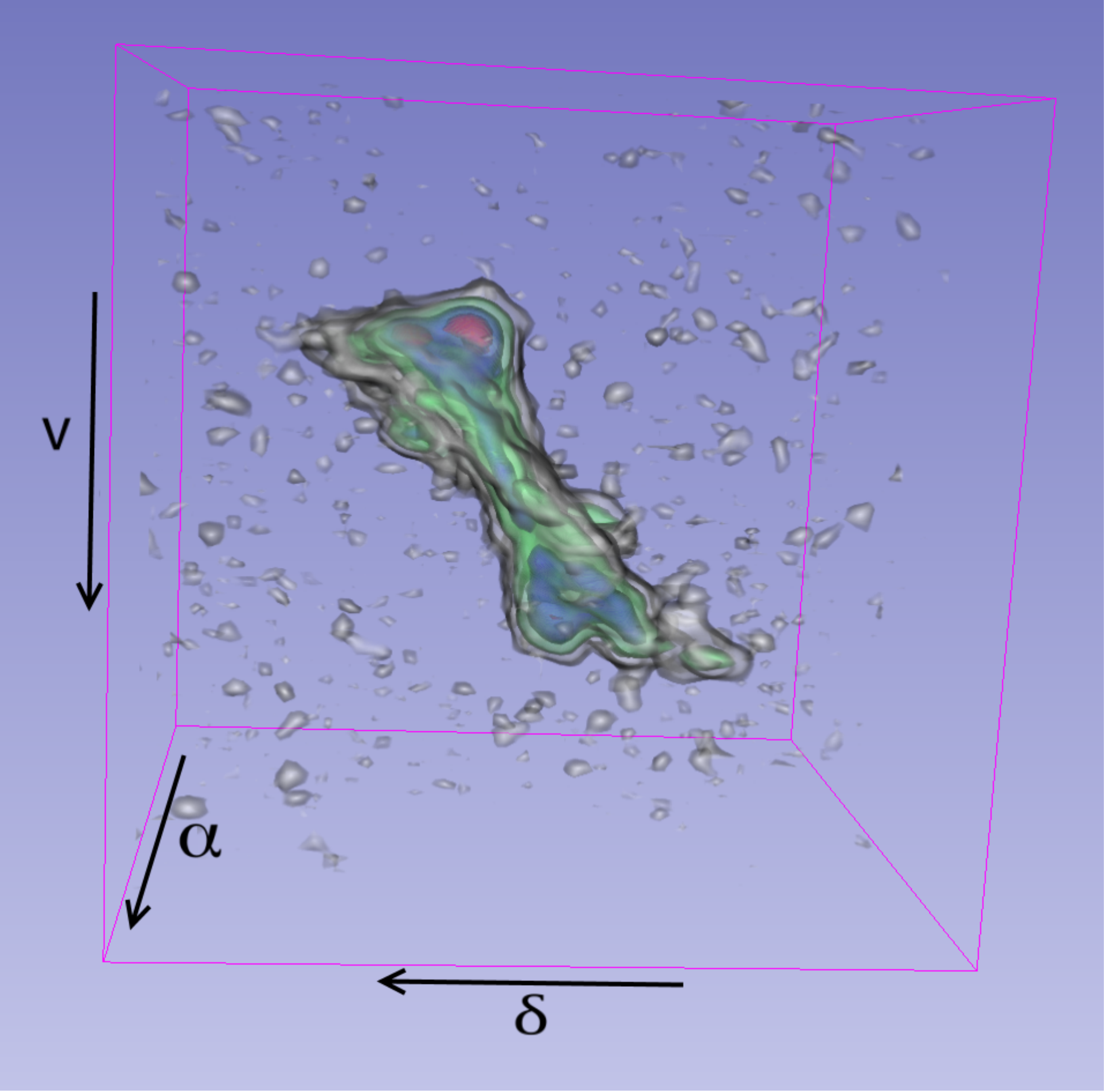}
\caption{Three views of the volume rendering of a particular source in the PPScl filament are 
shown. The optical counterpart, WEIN069, has been observed by \cite{wein}. The size of 
the data cube containing the source is $73^3 \sim 4 \times 10^5$ voxels. 
In the upper panel we look along the frequency axis; in the middle panel along the RA 
axis; and in the bottom the view is parallel to the geometrical major axis of the galaxy. 
For computing the projection we used an accumulate method. The different colors highlight 
different intensity levels in the data.}
\label{fig01}
\end{figure}

In this paragraph we will show in detail, using visualization examples, the character of the 
21-cm radio emission of galaxies and the benefits and drawbacks of adopting 3-D 
visualization, as pointed out already by \citet{Norris} and \citet{Oosterloo} (see also 
\cite{Goodman1}).  

The use of 3-D visualization of H\,{\small I} in galaxies is still in its infancy. Existing  
astronomical 3-D visualization tools lack interactivity and the ability to perform quantitative analysis. 
The lack of interactivity is mainly a result of the lack of computing power 
to date, as volume rendering is computationally expensive. Moreover, the use of 2-D 
input and output hardware limits the interaction with a 3-D representation  
(see section \ref{vireal}). Therefore, the interpretation of a 3-D visual 
representation has never been investigated thoroughly. Additional complication is that the 3-D structure 
of the H\,{\small I} in a cube is not in a 3-D spatial domain. The third axis represents velocity and 
thus the 3-D rendering delivers a mix of morphological, kinematical and geometrical information. Therefore, 
3-D visual analytics has never been developed for H\,{\small I} data. These are the main reasons why the 
development of 3-D visualization as a tool for inspecting, understanding and analysing 
radio-astronomical data has been slow. Currently available hardware, e.g. GPUs, now enable 
interactive volume rendering, stimulating further development.

3-D visualization techniques can provide many insights about the source under study. 
In Fig.\ref{fig01}, the three-dimensio-nal visualization of a particular source in the PPScl 
filament, discussed in section \ref{Large}, shows a 3-D view of its H\,{\small I} distribution and 
kinematics providing an immediate overview of the structures in the data. Two main components 
are visible in Fig.\ref{fig01}: a central body, which is the regularly rotating disk of the 
galaxy, and a tail which is unsettled gas resulting from tidal interaction with another galaxy. 
The 3-D structure of the H\,{\small I} data is, however, difficult to interpret for several reasons: i) 
the third axis of a data cube is frequency, which is converted into a velocity applying the 
Doppler formula to the 21-cm H\,{\small I} line; ii) the measured velocities are the line-of-sight 
velocity components of a rotating system, therefore the 3-D shape depends directly on the 
rotation curve; iii) in addition, the kinematic information of the gas is affected by geometric 
properties such as inclination, orientation of the semi-major axis, and gas distribution. Due to 
these complexities in the data, the user of a 3-D inspection tool needs reasonable experience 
with the data and a certain learning period to assimilate the tool itself. This is not different 
from the situation 25 years ago, when radio astronomers had to train themselves to understand 
2-D visual representations such as movies of channel maps and position-velocity 
diagrams. During this learning process interactivity is a key-factor (see \ref{finder} and 
\ref{analysis}). 

The 3-D visualization paradigm (volume rendering) described and used in this paper is limited by 
the use of 2-D input and output hardware such as a standard monitor and mouse. A simple practical 
example of a limitation in 3-D is the absence of a method for picking the value of one
pixel with a \textit{cursor}. 
Complementary visualization in 1-D and 2-D can repair these deficiencies. Moreover, there is not 
a single best way to visualize a radio data cube, but the combination of several
methods (3-D, 2-D, 1-D, side by side,
overlaid, blinking, etc.) and the interaction between them could deliver a very powerful 
analysis tool. It is important to view the data in different ways; this is the key to fully 
assimilating the information in the data. Therefore, a high-level of 1-D/2-D/3-D linked views 
must be achieved.

\begin{figure}[h!]
\centering
\includegraphics[width=0.48\textwidth]{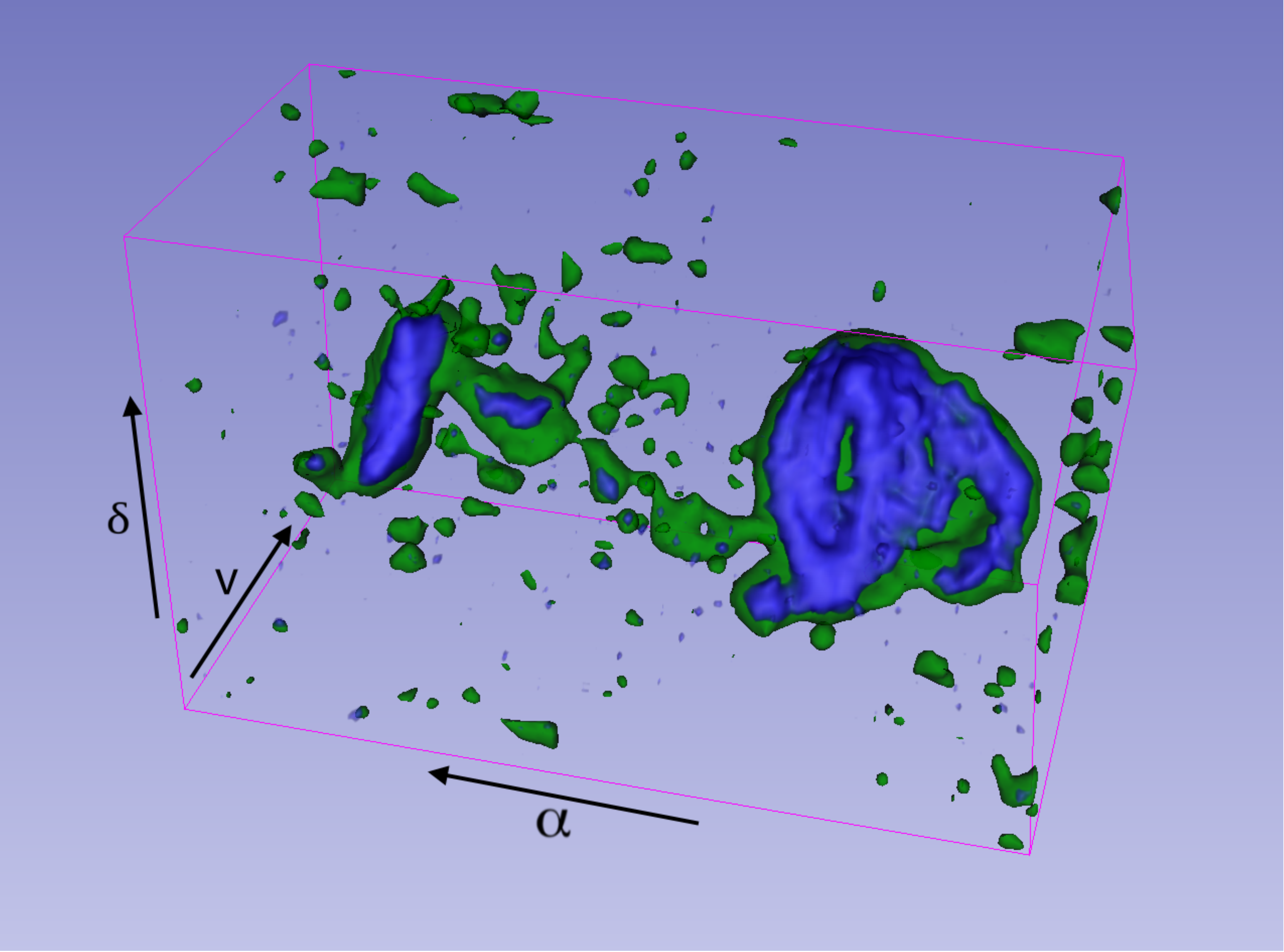}
\caption{Another view of the source in Fig.\ref{fig01} is shown. The blue surface represents the 
full resolution data, while the green is the smoothed version at 60" spatial 
resolution. Both surfaces are representations of the signal at $3\sigma$. The green surface 
shows a very faint filamentary structure that connects the two galaxies.}
\label{fig02}
\end{figure}

Very faint coherent signals, under $3\sigma$, are difficult to find even using 3-D. Real-time 
smoothing can help in dealing with the noisy character of the data. In fact, if the signal is 
comparable to the noise, which will be the case for many APERTIF observations, it is not 
possible to distinguish the signal itself from the noise at full resolution in any way. 
In Fig.\ref{fig02}, it is shown that only in the smoothed (60") version of the same 
data (in this case the signal-to-noise ratio of the filament is increased from $\sim 1$ to 
$\sim 4$) it is possible to localize a very faint filamentary structure that connects the two galaxies. 
It is is already possible to detect the filament after applying a smoothing to a 
spatial resolution of 30" (signal-to-noise of $\sim 2$).  

In the following use cases we will show how 3-D interactive visualization helps 
in the analysis of the sources.

\subsubsection{Use Case A: analysis of sources with tidal tails}\label{finder}
\begin{figure*}
\centering
\includegraphics[width=0.495\textwidth]{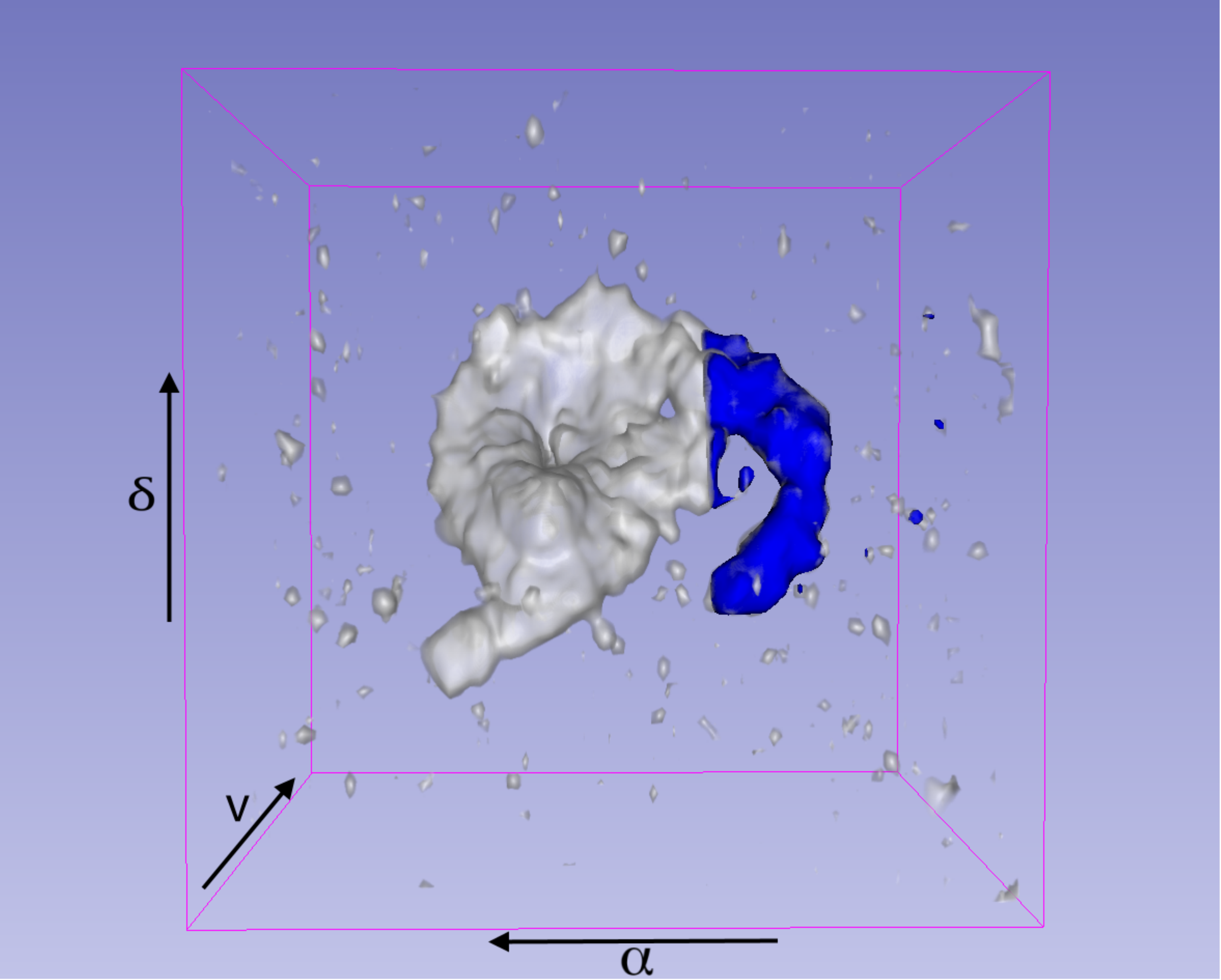}
\includegraphics[width=0.495\textwidth]{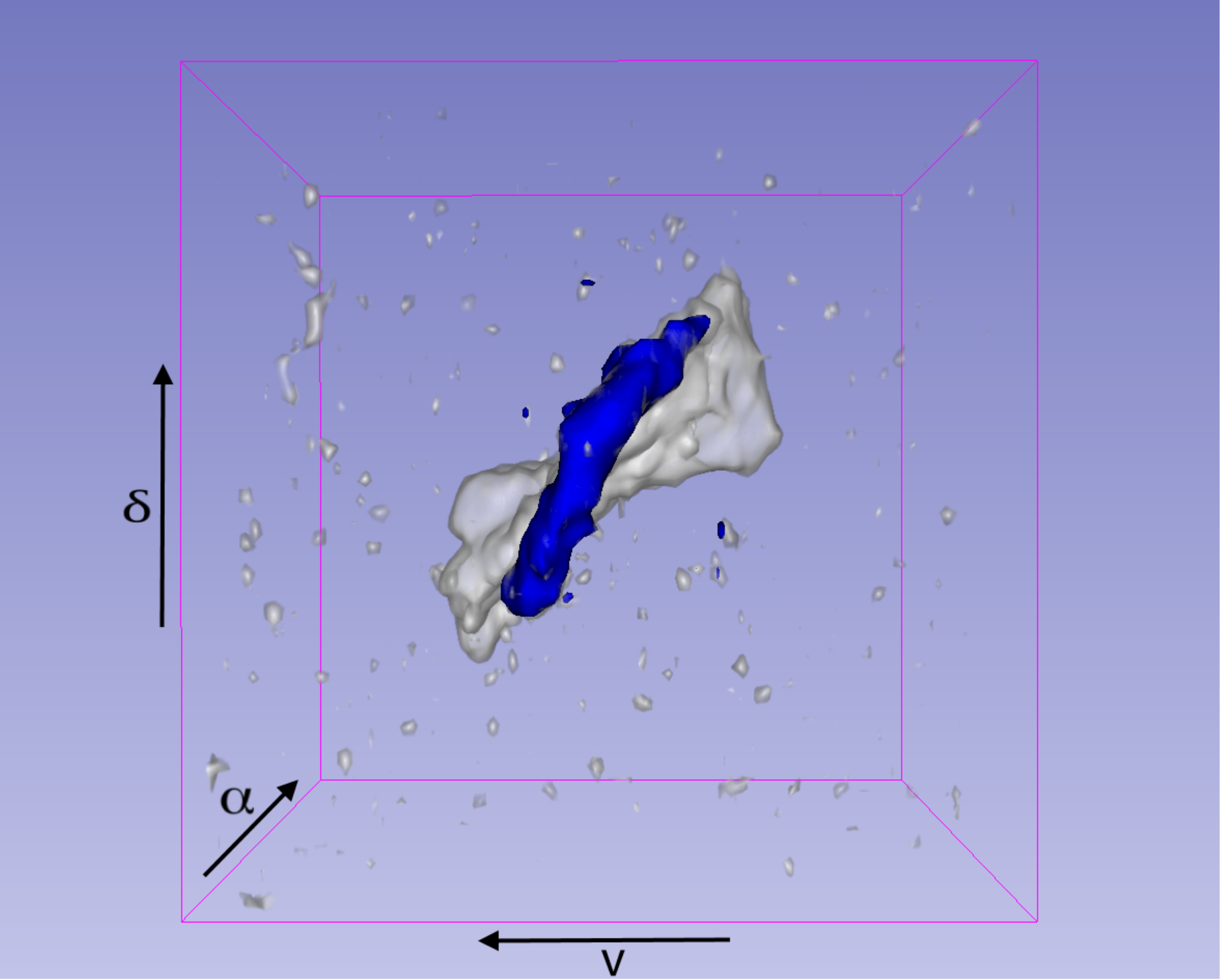}
\caption{Another two views of the source in Fig.\ref{fig01} are shown. The blue surface 
is a manual selection of the tidal tail. }
\label{fig1}
\end{figure*}

Fig.\ref{fig1} explores the source shown in Figs.\ref{fig01} and \ref{fig02} in more detail. A 
big tail due to a gravitational interaction is clearly present in the data cube. It is very easy 
to recognize the tail in the volume rendering because the data are coherent in all three 
dimensions. 

In the case of H\,{\small I} in galaxies one can extract additional information from fitting the 
observations with the so called \textit{tilted-ring} model \citep{Warner}. Modeling tools 
(e.g., $\tt{TiRiFic}$ \citep{Jozsa}; $\tt{^{\rm3D}\,Barolo}$ \citep[][in prep.]{Enrico}) generate a 
parametrized model data cube, simulating the observed H\,{\small I} distribution of the galaxy as a set 
of concentric, but mutually inclined, rotating rings, which is then compared directly 
to the observation. This operation can give a deeper knowledge of the kinematics and morphology: 
asymmetries in surface density and velocity, presence of extra-planar gas, presence of inflows 
and outflows, gas at anomalous velocities, etc. However, these algorithms cannot recognize 
tidal tail structures and separate them from the central regularly rotating body of the 
galaxy. Combining 3-D visualization with these algorithms through a 3-D selection tool 
\citep[e.g.][]{Yun} will be very powerful. As shown in Fig. \ref{fig1}, separating 
the components visually enables a better view and a better understanding compared to the visual 
representation shown in the middle panel in Fig.\ref{fig01}.

A 3-D selection tool will not only be useful for highlighting the different components with 
different colors, but also for retrieving quantitative information (noise calculation, H\,{\small I} 
mass, velocity gradient, tilted-ring model-fitting, etc.) on the selected volume. For example, 
in the case of this PPScl source the user can separate the components and perform the 
calculations separately on the two volumetric selections. In this process, the key-feature is 
the interactivity of the process itself.

\subsubsection{Use Case B: modeling feedback}\label{analysis}  

\begin{figure}
\centering
\includegraphics[width=0.48\textwidth]{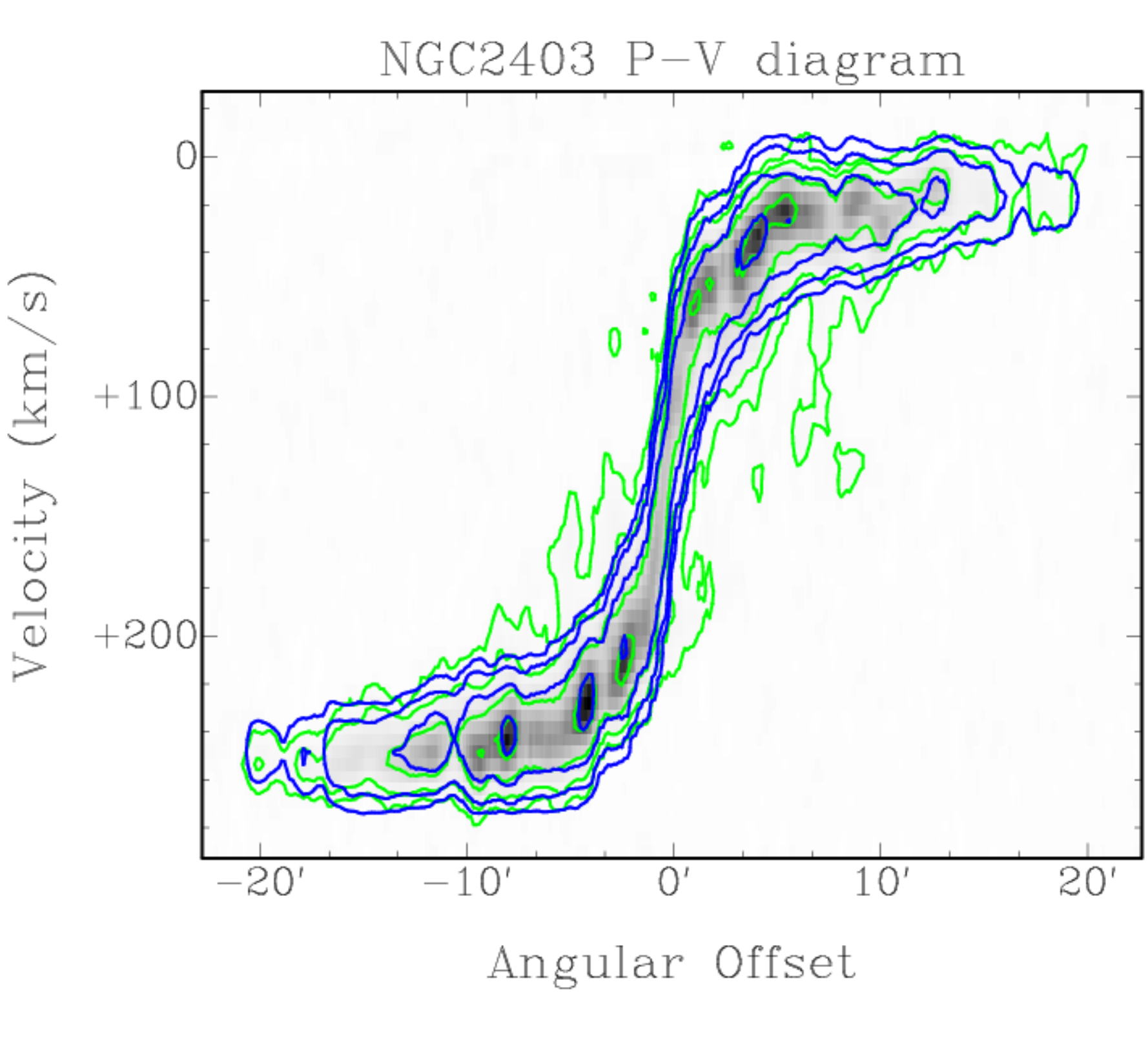}
\includegraphics[width=0.48\textwidth]{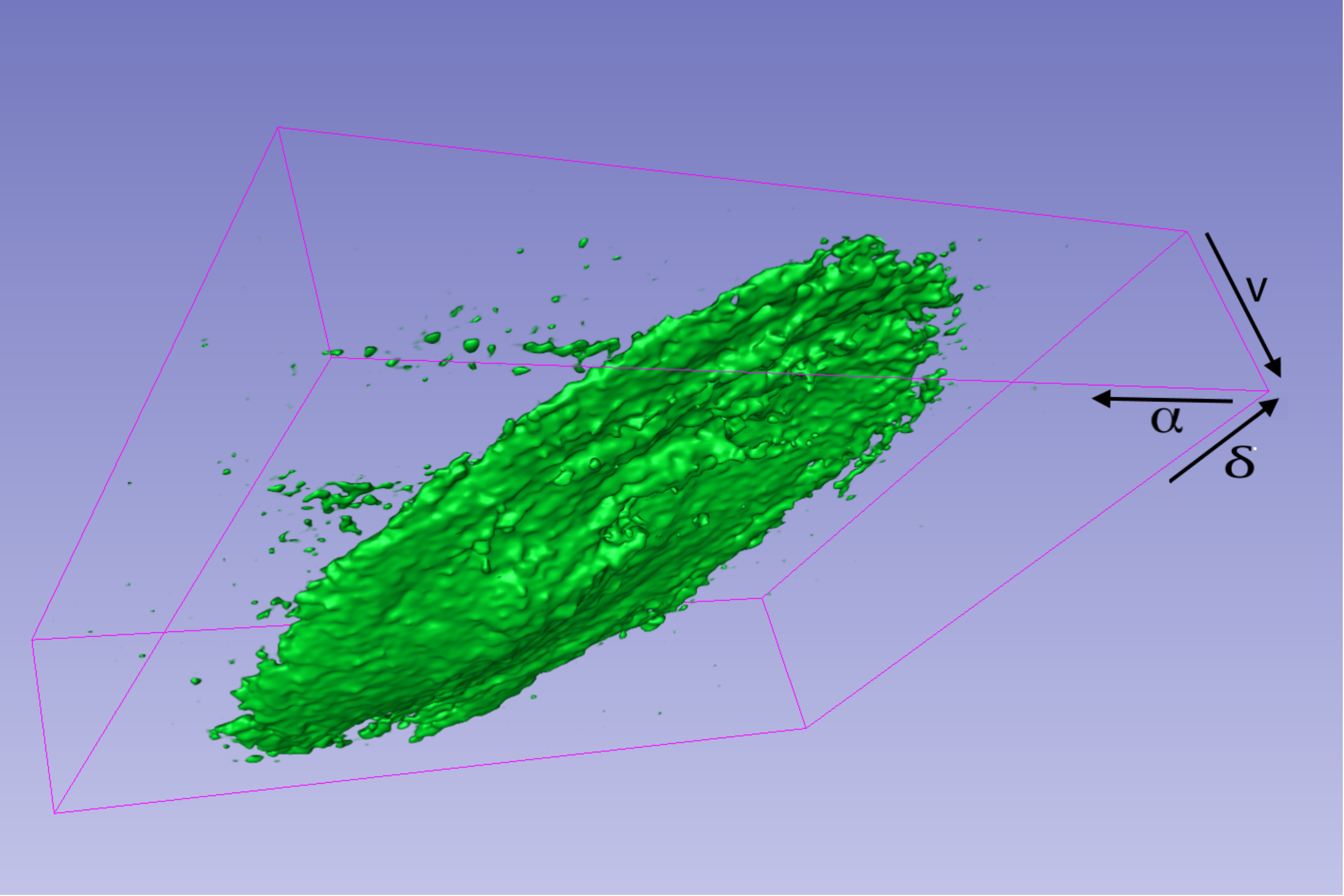}
\includegraphics[width=0.48\textwidth]{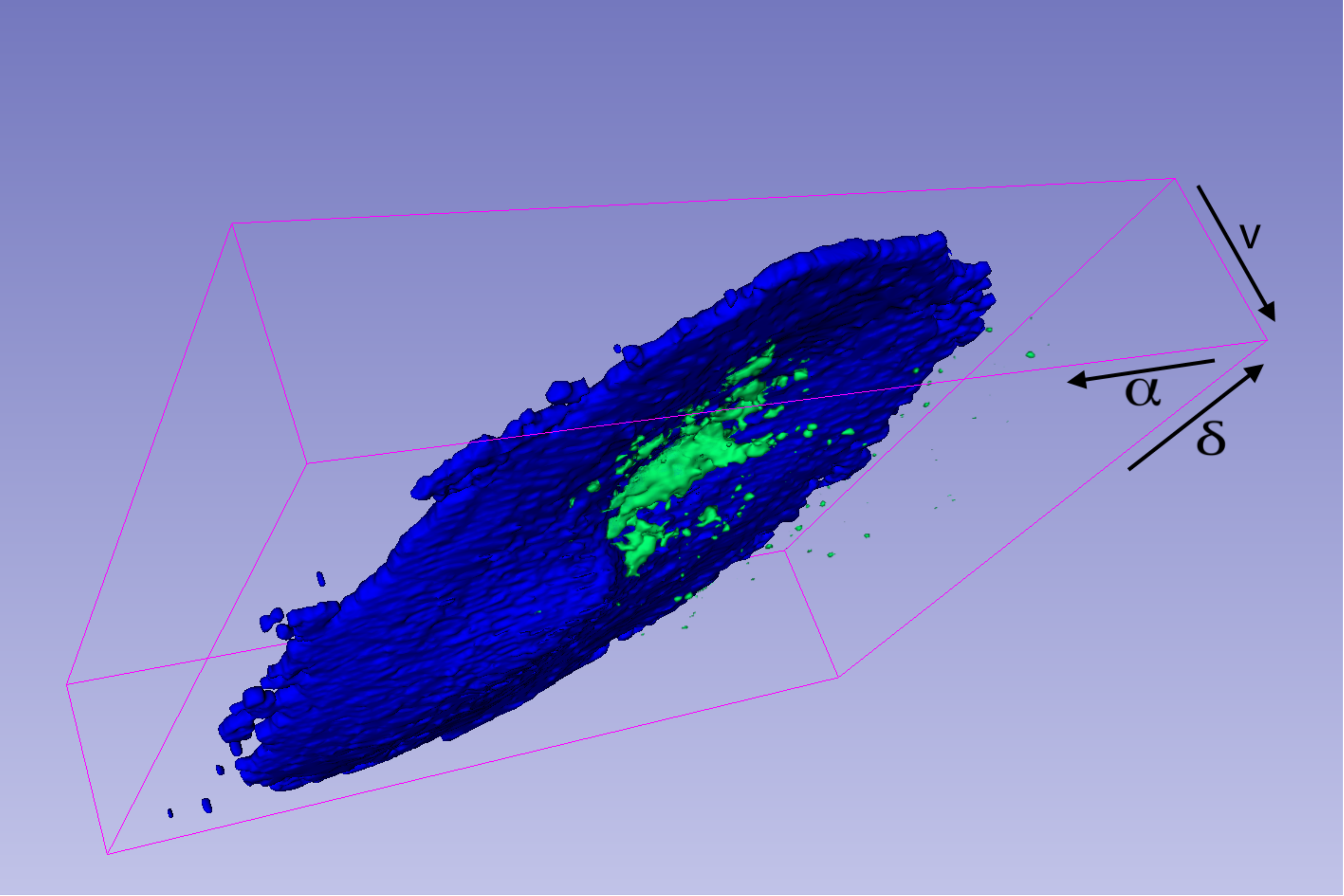}
\caption{Three different illustrations of the H\,{\small I} data of NGC2403 from the THINGS 
survey \citep{Walter} are shown. The galaxy is very well resolved. The top panel represents 
the position-velocity diagram along the semi-major axes which shows the typical rotation curve 
of a late-type galaxy (the blue contours represent the model that fits the regular disk) plus 
some unsettled gas in the inner region (the lowest green contour of the data is at $3\sigma$). 
The middle and bottom panels illustrate two 3-D representation of the data using an accumulate 
projection method. }
\label{fig2}
\end{figure}

It has been shown that the gas distribution of some spiral galaxies (e.g., NGC2403 shown in 
Fig.\ref{fig2}) is not composed of just a cold \textit{regular} thin disk. Stellar winds and 
supernovae can produce extra-planar gas (e.g., galactic fountain \citep{Bregman}). In this case, 
modelling can be used to constrain the 3-D structure and kinematics of the extra-planar gas 
which is visible in the data as a faint kinematic component in addition to the disk. 
3-D visualization of both the data and the model can provide a powerful tool to investigate such 
features. The visualization tool could use the output model of automated model-fitting 
algorithms for visually highlighting the different components in the data cube. In 
fact, if the model of the cold thin disk is subtracted from the data, it is possible immediately 
to locate any uncommon features in the data cube of interest and have already an idea of their 
properties, directing further modeling. For example, a model of the extra-planar gas above or 
below the disk with a slower rotation and a vertical motion provides quantitative information 
about the rotation and the infall velocity of such gas. 

In Fig.\ref{fig2}, the data of the NGC2403 observations are colored in green, while the blue 
structure is a tilted-ring model of regular rotation automatically fitted to the data with 
$\tt{^{\rm3D}\,Barolo}$. The top panel in Fig.\ref{fig2} represents the position-velocity 
diagram along the semi-major axes which shows the typical rotation curve of a late-type galaxy 
plus some unsettled gas in the inner region. The middle panel is a 3-D representation of the 
data, but it is very difficult to distinguish between the cold disk and the extra-planar gas. In 
fact, too much information is condensed in that visual representation. Separating and visually 
highlighting the different kinematic components, such as in the bottom panel, clearly shows the 
extra-planar gas. 3-D visualization gives an immediate overview of the coherence. For example, 
it highlights the presence of extra-planar gas and its extension. On the other hand, for 
checking the data pixel by pixel it is better to use a two-dimensional representation like a 
position-velocity diagram.

\section{Prerequisites for visualization of H\,{\small I} }\label{prere}

\cite{Goodman} has already expressed that a visualization environment for astronomy 
should satisfy:

\begin{enumerate}[i)]
\item interactivity;
\item linked views with different representations of the data (2-D, 3-D and high-dimensional 
visualization);
\item availability of an open source repository and a high level of modularity in the source 
code for enabling collaborative work; 
\item interoperability with Virtual Observatory (VO) tools through the Simple Application 
Messaging Protocol \citep[SAMP;][]{samp}.
\end{enumerate}

These requirements are also valid in our case, the visualization of H\,{\small I} in galaxies. 
Moreover, the interface must be able to handle astronomical world coordinates. 
This is of primary importance for many applications such as overlaying images taken at different 
wavelengths with other telescopes, cross-correlating source positions and velocities with existing 
catalogs, etc. A full overview of representation methodologies
of celestial coordinates in FITS and related issues is given in \cite{Calabretta1} and 
\cite{Calabretta2}.

From section \ref{Large} we concluded that the data cubes of 
interest will have dimension $< 10^7$ voxels ($< 0.25$ GB), but a large number ($\sim 
100,000$) of small sources will be delivered by the surveys. Therefore, for quickly extracting  
the information from the data and presenting them in a clear and synthetic form, the 
visualization must be qualitative, quantitative, and comparative. In the next three paragraphs 
we will describe these demands and why we need three \textit{levels} of visualization.

\subsection{Qualitative visualization}

First of all, astronomers want to look at their data in various ways in order to assess the data 
quality. An experienced astronomer can distinguish faint sources from the 
noise and instrumental artefacts, recognize the morphology and the kinematics of a galaxy, and 
identify unexpected H\,{\small I} emission (e.g., very faint structures such as extra-planar gas, tidal 
tails, and ram-pressure filaments). Therefore, qualitative visualization will continue to play a major 
role. 

In the previous section we showed the advantages and the drawbacks of adopting 3-D 
visualization. Very fast interactivity in rendering, in 3-D 
navigation, in data smoothing, and in quantitative and comparative functionality is important: if the 
interactivity is too slow, only the obvious signal will be found and subtle features may remain 
unnoticed. More precisely, the visualization should have a user-friendly interface capable to 
sustain navigation with more than 15 fps in order to provide the user with a fast interaction 
such as rotation, zooming, and panning of the data. 

The interface should have the capability to change the transfer function (i.e., mapping 
the value of the projected voxels onto a color and transparency value) interactively to help 
the astronomer in the qualitative understanding of the data, both in the 2-D and 3-D 
visualization.
 
The user should also be able to choose different line-of-sight integrations during 
the process of projection for the volume rendering (e.g., minimum, maximum, accumulate). 
For example, in order to visualize H\,{\small I} absorption that is a negative line, a minimum transfer 
function is needed, while to see the H\,{\small I} emission in galaxies one can use a maximum or a very 
specific accumulate transfer function.

\subsection{Quantitative visualization}
Interactive quantitative visualization which allows the user to extract quantitative information 
directly from the visual presentation is of primary importance. In astronomy, and in particular 
in radioastronomy, this is not a new concept. For example the $\tt{KARMA}$ package is a very 
good quantitative tool in the framework of 1-D and 2-D visualization. $\tt{KARMA}$ developers 
showed that a first level of quantification is to retrieve numbers from the visualized dataset 
and in some cases to represent them in a visual way for a better understanding. Examples are:

\begin{enumerate}[i)]
 \item display of the flux value through a pixel in slice view and/or plot intensity 
 profiles and display the value;  
 
 \item calculation of noise, standard deviation, maximum, minimum, H\,{\small I} mass or velocity 
 gradient, etc., in a specific area or volume;
 
 \item segmentation of the 3-D data volume of an object; 
 
 \item construction and display of moment maps and position-velocity diagrams. 
  
\end{enumerate}

A second level of quantification can be introduced by having interactive features between the 
visualization and a plotting library (see, for example, the work in progress by \cite{Goodman} 
and her team related to the $\tt{GLUE}$ Project 
\footnote{\url{http://projects.iq.harvard.edu/seamlessastronomy/software/glue}}). The idea is to 
plot quantitative information related to the data and then have a visual representation of that 
information in the visualization of the data.
In order to give an idea of the benefits of this functionality, a hypothetical example 
follows: the first step is downloading H\,{\small I}, optical and infrared data, creating star formation 
rate (SFR) maps and plotting the local SFR values as a function of the H\,{\small I} column density 
(N$_{\rm H\,{\small I}}$) of the correspondent pixel. The plot allows the identification of pixels 
deviating from the power law relation between SFR and N$_{\rm H\,{\small I}}$. Subsequently, it will be 
possible to locate possibly deviant pixels by highlighting them in the 3-D 
visualization. The second step is to examine where they are in the 3-D data in order to assess whether 
they occupy specific regions, i.e., if they are coherent in the 3-D data. The third step is 
retrieving quantitatively the SFR of a specific environment of the data cube under investigation. 
For that it is necessary to select different zones using the visualization and then to plot 
the SFR/N$_{\rm H\,{\small I}}$ of each zone with a different color (for example two regions in a spiral 
galaxy: the spiral arms and the bulge).

Standalone quantitative visualization is however not satisfactory. A synergy, using linked 
views, with comparative visualization is necessary for assessing the quality of the analysis, 
such as comparing a tilted-ring model with the data, and highlighting subtle faint structure in 
the data as we have shown in section \ref{analysis}. 

\subsection{Comparative visualization}

In sections \ref{finder} and \ref{analysis} we showed how in the case of H\,{\small I} in galaxies one 
can extract additional information from tilted-ring model-fitting. 

The visualization tool should also enable an interactive comparison between data and models in 
order to check the quality of the model provided by the automated algorithm. This is possible 
by having the model routine embedded in the visualization interface. In fact, a coupling between 
model fitting and visualization will enable an interactive change of the parameters of the 
model, such as rotation curve, density column, and inclinations as function of the radius, and 
the comparison of the new model with the data. Interactive tilted-ring model fitting 
greatly helps in the analysis of warped galaxies. For example, \cite{sparke} adopted an interactive 
procedure, using $\tt{INSPECTOR}$, for arriving at the final model of NGC 3718 shown in the paper.
$\tt{INSPECTOR}$ is an interactive tilted-ring modeling routine in $\tt{GIPSY}$ using a comparative 
visualization tool. 

The comparison between an observation and a model of a galaxy can be made by examining 3-D 
renderings of the data and the model in two separate windows, or by showing in one window an overlay of the 
model on the observation and in another window the difference between them. This separates regularly 
rotating gas from \textit{unusual} kinematic features (extra planar gas, tidal tails, ram pressure induced 
structures). In addition,  the interface needs to support display windows next to the 3-D rendering with plots in 
which one can view results of the source analysis such as the rotation curve. 

Comparative visualization can be also extended using models obtained by running $N$-body simulations  
\citep[see][]{Identikit1,Identikit2}. This kind of systematic studies can benefit, in terms of 
speed and interactivity, from the usage of optimized $N$-body codes running on GPUs 
\citep{nyland,simon, HiGPUs}, some of which are publicly available via the Astronomical 
Multipurpose Software Environment \citep[AMUSE;][]{amuse}.

\subsection{High-dimensional visualization techniques}\label{highvisu}

High-dimensional data visualization (e.g. $\tt{TOPCAT}$ \citep{Topcat}) of the 
parameter tables will enable the capability to have a full picture of the characteristics 
of the data in the catalog.  This feature is very important to discover the unexpected. In 
fact, the catalog paradigm can fail if the number of sources is too large: in general 
it is possible to retrieve a list of data from catalogs using flags such as 
names or certain parameters of the objects; it is, however, usually not possible to have a general view 
of the main parameters of the sources in question. Therefore, a visualization package should be 
able to download tables that contain the required properties of galaxies  (flux, flux error, degree 
of asymmetry, velocity width, integrated profile shape, etc.) and plot these parameters, 
allowing the user to find outliers. The user should also have the capability to mark the data 
of interest in the plot and download the requested data cube(s) from the 
catalog, using the interface for further exploration of the 3-D signatures and comparing them 
with one or more models. This can be achieved using the SAMP protocol and other VO tools.

\subsection{Summary}\label{summeri}

In this section we have defined the requirements that visualization of H\,{\small I} emission, in the survey 
era, must satisfy. We briefly summarize them here:

\begin{enumerate}[a)]
\item astronomical world coordinates in order to combine the visualization of H\,{\small I} data with data obtained 
at other wavebands;
\item 3-D capabilities (i.e., presence of interactive volume rendering for grid data of dimension 
$ < 10^7$ voxels and interactive color and opacity function widgets);
\item interactive linking between 1-D/2-D/3-D views;
\item quantification: physical data units, labels, and statistical tools;
\item linked 1-D/2-D/3-D selection tools;
\item 3-D segmentation techniques;
\item interactive smoothing;
\item comparative visualization (multiple views, overlaid visualizations, etc.);
\item tools for generating tilted-ring and $N$-body models;
\item interoperability with VO tools.

\end{enumerate}

\section{Review of state of-the-art 3-D visualization packages} \label{review_sof}

In the previous section we described in detail all the requirements a visualization tool  
must satisfy for enabling the source analysis that we outlined in section \ref{3-D}.
A review of the current state-of-the-art of 3-D visualization is very important in order
to avoid duplication and development of rendering algorithms and tools which may already exist. 
We performed a review of current 3-D visualization software with the idea in mind that they have 
to satisfy the requirements listed in section \ref{summeri}, plus the following technical 
prerequisites. The software must:

\begin{enumerate}[i)]
\item run on multiple platforms;

\item have an intuitive interface;

\item have a Python wrapper for easy introduction of the SAMP protocol;

\item have a high level of modularity in the source code;

\item have proper documentation and long-term maintainability (i.e., presence of a significant user- and developer-community).
\end{enumerate}

Many rendering algorithms and tools exist but we restricted the detailed review to a short list 
of publicly available, open-source and currently maintained packages with 3-D interactive 
rendering capabilities:  

\begin{enumerate}[1)]
\item $\tt{Paraview}$ \citep{ParaView}: a general-purpose multi-platform data analysis and 
visualization application. The ParaView project started in 2000 as a collaborative effort 
between Kitware Inc. and Los Alamos National Laboratory.

\item $\tt{3DSlicer}$ \citep{Slicer}: a software package for visualization and image analysis of 
medical data. It is natively designed to be available on multiple platforms.

\item $\tt{Mayavi2}$ \citep{mayavi1,mayavi2}: a general purpose, cross-platform tool for 
2-D and 3-D scientific data visualization.

\item $\tt{ImageVis3D}$ \citep{img3dvis}: a new volume rendering program developed by the 
NIH/NIGMS Center for Integrative Biomedical Computing (CIBC). The software is multi-platform 
and scalable.

\end{enumerate}

For each package we performed a detailed review study in two steps:

\begin{enumerate}[i)]
\item a software user-friendliness survey: we tested the four packages by inspecting and 
analysing the H\,{\small I} emission of WEIN069 and NGC2403 (shown in Fig. \ref{fig01} and \ref{fig2}). 
We performed a survey by asking 15 radioastronomers to evaluate the 
intuitiveness and interactivity of the different features offered by each package using WEIN069 
as test data set. The evaluation involved each participant filling out a questionnaire after one hour of 
utilization of the packages. In all cases the latest stable version of the software was used  
with the following hardware set-up: a Linux laptop (Ubuntu 14.04 LTS) equipped with an Intel i7 
2.60 GHz CPU, an NVIDIA GeForce GTX860M GPU, 16 GB of DDR3 1.6GHz RAM, and
a 15.6 inch monitor with a resolution of $1920 \times 1080$.  
\item a source code evaluation: we performed a detailed study of the full source code, the 
level of modularity, and the available documentation for developers.
\end{enumerate}

\subsection{Review results}\label{list}
The resulting ranking of the packages is shown in Table \ref{tab}. In addition we 
provide a detailed list of pro's and con's for each package in Table \ref{listTab}.

\begin{table*}
\centering

\begin{tabular}{c|m{2.1cm}|m{2.1cm}|m{2.1cm}|m{2.1cm}|c}
\hline 
\hline
 &  $\tt{Paraview}$ & $\tt{3DSlicer}$ & $\tt{Mayavi2}$ & $\tt{ImageVis3D}$ & Description\\
\hline 
\hline
\textbf{a} & \vspace{0.2cm}\huge{\hspace{0.75cm}${\otimes}$}\vspace{0.1cm} & \vspace{0.2cm}\huge{\hspace{0.75cm}${\otimes}$}\vspace{0.1cm} & \vspace{0.2cm}\huge{\hspace{0.75cm}${\otimes}$}\vspace{0.1cm} & \vspace{0.2cm}\huge{\hspace{0.75cm}${\otimes}$}\vspace{0.1cm} & astronomical world coordinates\\
\hline
\textbf{b} 
&
\vspace{0.1cm}
\includegraphics[width=2.1cm, height=0.6cm]{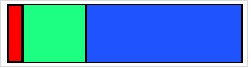}
\small{$7\% \,| \, 27\% \,| \, 67\%$}
&
\vspace{0.1cm}
\includegraphics[width=2.1cm, height=0.6cm]{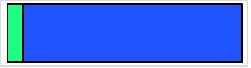}
\small{$0\% \,| \, 7\% \,| \, 93\%$}
&
\vspace{0.1cm}
\includegraphics[width=2.1cm, height=0.6cm]{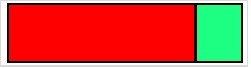}
\small{$80\% \,| \, 20\% \,| \, 0\%$}
&
\vspace{0.1cm}
\includegraphics[width=2.1cm, height=0.6cm]{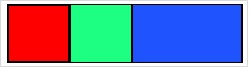}
\small{$27\% \,| \, 27\% \,| \, 47\%$}
&
3-D capabilities and color transfer function\\
\hline
\textbf{c}
&
\vspace{0.1cm}
\includegraphics[width=2.1cm, height=0.6cm]{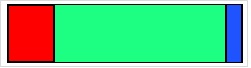}
\small{$20\% \,| \, 73\% \,| \, 7\%$}
&
\vspace{0.1cm}
\includegraphics[width=2.1cm, height=0.6cm]{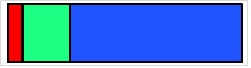}
\small{$7\% \,| \, 20\% \,| \, 73\%$}
&
\vspace{0.1cm}
\includegraphics[width=2.1cm, height=0.6cm]{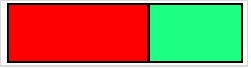}
\small{$60\% \,| \, 40\% \,| \, 0\%$}
&
\vspace{0.1cm}
\huge{\hspace{0.75cm}${\otimes}$}
&
linked 1-D/2-D/3-D views\\
\hline
\textbf{d}
&
\vspace{0.1cm}
\includegraphics[width=2.1cm, height=0.6cm]{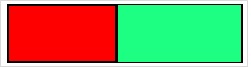}
\small{$47\% \,| \, 53\% \,| \, 0\%$}
&
\vspace{0.1cm}
\includegraphics[width=2.1cm, height=0.6cm]{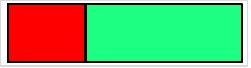}
\small{$33\% \,| \, 67\% \,| \, 0\%$}
&
\vspace{0.1cm}
\includegraphics[width=2.1cm, height=0.6cm]{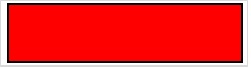}
\small{$100\% \,| \, 0\% \,| \, 0\%$}
&
\vspace{0.1cm}
\huge{\hspace{0.75cm}${\otimes}$}
&
data probe, labels and statistics \\
\hline
\textbf{e} 
&
\vspace{0.1cm}
\includegraphics[width=2.1cm, height=0.6cm]{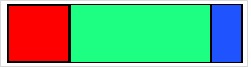}
\small{$27\% \,| \, 60\% \,| \, 13\%$}
&
\vspace{0.1cm}
\includegraphics[width=2.1cm, height=0.6cm]{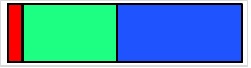}
\small{$7\% \,| \, 40\% \,| \, 53\%$}
&
\vspace{0.1cm}
\includegraphics[width=2.1cm, height=0.6cm]{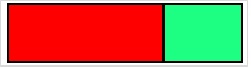}
\small{$67\% \,| \, 33\% \,| \, 0\%$}
&
\vspace{0.1cm}
\huge{\hspace{0.75cm}${\otimes}$}
&
linked 1-D/2-D/3-D selection tools\\
\hline
\textbf{f} 
&
\vspace{0.1cm}
\includegraphics[width=2.1cm, height=0.6cm]{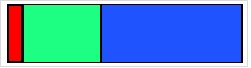}
\small{$7\% \,| \, 20\% \,| \,73\%$}
&
\vspace{0.1cm}
\includegraphics[width=2.1cm, height=0.6cm]{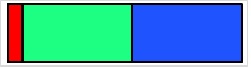}
\small{$13\% \,| \, 20\% \,| \, 67\%$}
&
\vspace{0.1cm}
\includegraphics[width=2.1cm, height=0.6cm]{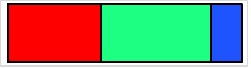}
\small{$47\% \,| \, 27\% \,| \, 27\%$}
&
\vspace{0.1cm}
\includegraphics[width=2.1cm, height=0.6cm]{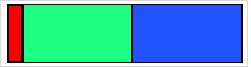}
\small{$7\% \,| \, 47\% \,| \, 47\%$}
& 3-D segmentation\\
\hline
\textbf{g} & \vspace{0.2cm}\huge{\hspace{0.75cm}${\otimes}$}\vspace{0.1cm} & \vspace{0.2cm}\huge{\hspace{0.75cm}${\otimes}$}\vspace{0.1cm} & \vspace{0.2cm}\huge{\hspace{0.75cm}${\otimes}$}\vspace{0.1cm} & \vspace{0.2cm}\huge{\hspace{0.75cm}${\otimes}$}\vspace{0.1cm} & interactive smoothing\\
\hline
\textbf{h}
&
\vspace{0.1cm}
\includegraphics[width=2.1cm, height=0.6cm]{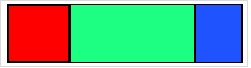}
\small{$27\% \,| \, 53\% \,| \,20\%$}
&
\vspace{0.1cm}
\includegraphics[width=2.1cm, height=0.6cm]{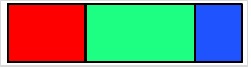}
\small{$33\% \,| \, 47\% \,| \, 20\%$}
&

\vspace{0.1cm}\huge{\hspace{0.75cm}${\otimes}$} & \vspace{0.1cm}\huge{\hspace{0.75cm}${\otimes}$} & comparative views \\
\hline
\textbf{i} & \vspace{0.2cm}\huge{\hspace{0.75cm}${\otimes}$}\vspace{0.1cm} & \vspace{0.2cm}\huge{\hspace{0.75cm}${\otimes}$}\vspace{0.1cm} & \vspace{0.2cm}\huge{\hspace{0.75cm}${\otimes}$}\vspace{0.1cm} & \vspace{0.2cm}\huge{\hspace{0.75cm}${\otimes}$}\vspace{0.1cm} & tilted-ring/$N$-body models routines \\
\hline
\textbf{j} & \vspace{0.2cm}\huge{\hspace{0.75cm}${\otimes}$}\vspace{0.1cm} & \vspace{0.2cm}\huge{\hspace{0.75cm}${\otimes}$}\vspace{0.1cm} & \vspace{0.2cm}\huge{\hspace{0.75cm}${\otimes}$}\vspace{0.1cm} & \vspace{0.2cm}\huge{\hspace{0.75cm}${\otimes}$}\vspace{0.1cm} & SAMP and VO connectivity\\
\hline 
\hline
\textbf{i} &  \vspace{0.2cm}\huge{\hspace{0.75cm}$\color{blue}\blacksquare$} & \vspace{0.2cm}\huge{\hspace{0.75cm}$\color{blue}\blacksquare$} & \vspace{0.2cm}\huge{\hspace{0.75cm}$\color{blue}\blacksquare$} & \vspace{0.2cm}\huge{\hspace{0.75cm}$\color{blue}\blacksquare$} & multi platforms software\\
\hline
\textbf{ii}
&
\vspace{0.1cm}
\includegraphics[width=2.1cm, height=0.6cm]{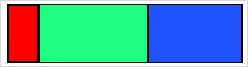}
\small{$13\% \,| \, 47\% \,| \, 40\%$}
&
\vspace{0.1cm}
\includegraphics[width=2.1cm, height=0.6cm]{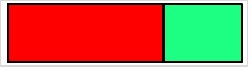}
\small{$67\% \,| \, 33\% \,| \, 0\%$}
&
\vspace{0.1cm}
\includegraphics[width=2.1cm, height=0.6cm]{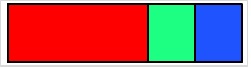}
\small{$60\% \,| \, 20\% \,| \, 20\%$}
&
\vspace{0.1cm}
\includegraphics[width=2.1cm, height=0.6cm]{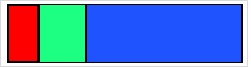}
\small{$13\% \,| \, 20\% \,| \, 67\%$}
&
intuitiveness of the interface\\
\hline
\textbf{iii} & \vspace{0.2cm}\huge{\hspace{0.75cm}$\color{blue}\blacksquare$} & \vspace{0.2cm}\huge{\hspace{0.75cm}$\color{blue}\blacksquare$} & \vspace{0.2cm}\huge{\hspace{0.75cm}$\color{blue}\blacksquare$} & \vspace{0.2cm}\huge{\hspace{0.75cm}$\color{red}\blacksquare$} & Python wrapper\\
\hline
\textbf{iv} & \vspace{0.2cm}\huge{\hspace{0.75cm}$\color{blue}\blacksquare$} & \vspace{0.2cm}\huge{\hspace{0.75cm}$\color{blue}\blacksquare$} & \vspace{0.2cm}\huge{\hspace{0.75cm}$\color{green}\blacksquare$} & \vspace{0.2cm}\huge{\hspace{0.75cm}$\color{red}\blacksquare$} & modularity of the software\\
\hline
\textbf{v} & \vspace{0.2cm}\huge{\hspace{0.75cm}$\color{blue}\blacksquare$} & \vspace{0.2cm}\huge{\hspace{0.75cm}$\color{blue}\blacksquare$} & \vspace{0.2cm}\huge{\hspace{0.75cm}$\color{green}\blacksquare$} & \vspace{0.2cm}\huge{\hspace{0.75cm}$\color{red}\blacksquare$} & documentation/long-term maintainability\\
\hline 
\hline
\end{tabular}
\newline
\newline

\begin{tabular}{m{2cm}|m{3cm}|m{4.8cm}|m{3.7cm}|m{2.5cm}}
\hline
\vspace{0.1cm} \Large Legend: & \vspace{0.1cm} \Large $\otimes =$ missing  & \vspace{0.1cm} \Large $\color{red}\blacksquare$ $=$ not satisfactory & \vspace{0.1cm} \Large $\color{green}\blacksquare$ $=$ satisfactory & \vspace{0.1cm} \Large $\color{blue}\blacksquare$ $=$ good \tabularnewline
\hline
\end{tabular}
\caption{A ranking of several 3-D visualization packages is shown. In the top 
part of the table, the letters in the first column refer to the summarized requirements in 
section \ref{summeri}. In the bottom part, the roman numerals refers to the technical 
prerequisites listed in this section \ref{review_sof}. The colored bars are a 
representation of the ranking based on a user-test survey performed with 15 
radioastronomers. Note that this software ranking is oriented towards the visualization of 
H\,{\small I} data (grid volume of dimension $< 10^7$ voxels) in a desktop environment. }
\label{tab}
\end{table*}

\begin{table*}
  \centering
  \begin{tabular}{ | c | m{8.5cm} | m{6.5cm} | }
    \hline
    \textbf{Software} & \hspace{3cm} \textbf{Pro's} & \hspace{3cm} \textbf{Con's} \\ \hline
    
    $\tt{Paraview}$
   
    &
    
    \begin{itemize}
    \setlength\itemsep{-0.45em}
    \item CPU/GPU rendering based on the Visualization Toolkit (VTK);
    \item skill to connect to a server to do the computation; 
    \item editable interface with unlimited 2-D/3-D views;
    \item linked 1-D/2-D/3-D views;
    \item cropping and selection tools;
    \item 3-D segmentation techniques, i.e., isosurfaces;
    \item skill to perform statistics on the user selection;
    \item high level of modularity in the source code;
    \item embedded python console in the interface for fast interaction with the source code;
    \item presence of documentation both for users and developers. 
    \end{itemize}

    & 
    
    \begin{itemize}
    \setlength\itemsep{-0.45em}
    \item the interface is complex;
    \item astronomical world coordinates and labels not displayable;
    \item the interface is not optimized for 1-D and 2-D visualization;
    \item interactive smoothing missing.
    \end{itemize}
    
    \\ \hline
    
    $\tt{3DSlicer}$
    
    &
    
    \begin{itemize}
    \setlength\itemsep{-0.45em}
    \item CPU/GPU rendering based on VTK;
    \item interface is also optimized for 2-D visualization of channel maps;
    \item high-level of linking between 2-D and 3-D views;
    \item interactive cropping and selection editor tools;
    \item skill to perform statistics on the user selection;
    \item 3-D segmentation techniques, i.e., isosurfaces; 
    \item high level of modularity in the source code;
    \item embedded python console in the interface for fast 
    interaction with the source code;
    \item presence of documentation both for users and developers.
    \end{itemize}
    
    &
    
    \begin{itemize}
    \setlength\itemsep{-0.45em}
    \item the interface is very complex and not intuitive;
    \item astronomical world coordinates and labels not displayable;
    \item 1-D visualization missing;
    \item 2-D contour plots missing;
    \item interactive smoothing missing.
    \end{itemize}
    
    \\ \hline
    
    $\tt{Mayavi2}$
    
    &
    
    \begin{itemize}
    \setlength\itemsep{-0.45em}
    \item CPU rendering based on TVTK (Python wrapper for VTK);
    \item cropping and selection tools;
    \item 3-D segmentation techniques, i.e., isosurfaces; 
    \item contour plots;
    \item a simple and clean scripting interface in Python, 
    easy integration with other python libraries.
    \end{itemize}
    
    &
   
    \begin{itemize}
    \setlength\itemsep{-0.45em}
    \item the interface is not stable;
    \item presence of only CPU rendering capabilities. The frame rate per 
    second is low, $fps < 5$, for data cubes bigger than $10^6$ voxels;
    \item color transfer function widget is not easy to use;
    \item astronomical world coordinates and labels not displayable;
    \item 1-D visualization missing;
    \item interactive smoothing missing;
    \item lack of statistics tools.
    \end{itemize}
    
    \\ \hline
    
    $\tt{ImageVis3D}$
    
    &
    
    \begin{itemize}
    \setlength\itemsep{-0.45em}
    \item very light, fast, and intuitive interface;
    \item GPU rendering;
    \item 3-D segmentation techniques, i.e., isosurfaces.
    \end{itemize}
    
    &
    
    \begin{itemize}
    \setlength\itemsep{-0.45em}
    \item the long-term maintainability of the rendering code is unknown;
    \item astronomical world coordinates and labels not displayable;
    \item 1-D and 2-D visualization missing;
    \item interactive smoothing missing;
    \item lack of statistics tools;
    \item lack of documentation.
    \end{itemize}
    
    \\ \hline
  \end{tabular}
  \caption{A list of pro's and con's relative to the four packages is 
  presented. The advantages and disadvantages listed form a detailed 
  description of the feedback provided 
  by the authors and the users of the software survey shown in 
  Tab.\ref{tab}. }\label{listTab}
\end{table*}

We can divide the packages in two classes: i) $\tt{Paraview}$ and $\tt{3DSlicer}$; ii)  
$\tt{Mayavi2}$ and $\tt{ImageVis3D}$. The software in the first group has many 
features, while the second group mainly offers qualitative visualization. The users 
noted that the interfaces offered by $\tt{Paraview}$ and $\tt{3DSlicer}$ are complex, but at 
the same time, most of the users found $\tt{Paraview}$ rather intuitive. The intuitiveness (i.e., 
the learning time) ranking shown in Table \ref{tab} obviously also depends on the experience of the users 
with similar visualization software.

The review highlighted that the users experienced a major lack of functionality 
in all four packages for: 
displaying labels with proper astronomical coordinates; 1-D visualization (e.g. line profiles); 
interactive smoothing; simple editing or blanking, and specific operations such as constructing a 
position-velocity diagram along a specified spatial axis; and comparative visualization (e.g., overlaid 
1-D profiles and overlaid 2-D contour plots on another image). This is not a surprising result. 
In fact, the packages considered in this section are aimed towards general or medical visualization 
purposes and lack the specialized visualization representations and interaction aspects 
common in radio astronomy. On the other hand, they do have advanced rendering capabilities, such 
as provided by the Visualization Toolkit \footnote{\url{http://www.vtk.org/}} (VTK), and a modern, 
multiple-platform, reliable interface based on Qt \footnote{\url{http://qt-project.org/}}. 
For example, the packages enable the user to save the whole working session in a bundle: the 
data, the visualization, and the module structure used for the analysis. 

At at the moment, 
all the packages listed lack multi-volume rendering. Multi-volume rendering is the operation 
to render two or more volumes on the same space. This feature is necessary for enabling very 
fast 3-D overlaid comparative visualization.

\subsection{Visualization of H\,{\small I} and $\tt{3DSlicer}$}\label{sli}

Despite the complexity of the interface, we chose to adopt $\tt{3DSlicer}$ as base 
platform for the development of a H\,{\small I} visualization tool. Our choice has been the result of 
considering various factors such as the presence of adequate documentation, the number of people 
actively working on the software, and quantitative features already implemented in the interface. 
These three main factors make $\tt{3DSlicer}$ the best solution for us. In fact, the medical 
visualization needs are indeed very close to the astronomical ones. For example, the  
interface layout and the navigation through the data are already optimized for parallel 2-D visualizations 
(e.g., movies of channel maps). The following features need to be added to $\tt{3DSlicer}$ in order
to fulfill the requirements described in section \ref{prere}:

\begin{enumerate}[i)]

\item proper visualization of astronomical data cubes using the data formats FITS, HDF5, CASA, and Miriad;

\item enabling interactive smoothing in all three dimensions and multi-scale analysis, such as wavelet lifting;

\item generation of flux density profiles, moment maps and position-velocity diagrams 
linked with the 3-D view;

\item interactive 3-D selection of H\,{\small I} sources;

\item interactive H\,{\small I} data modeling coupled to visualization;

\item introduction of the SAMP protocol to enable interoperability with $\tt{Topcat}$, 
and other VO tools and catalogs.

\end{enumerate}

\section{Concluding Remarks}

H\,{\small I} observations are moving into the era of big surveys. Upcoming H\,{\small I} 
surveys, such as those envisaged with APERTIF and ASKAP, will deliver big data sets 
leading the radio astronomer into the regime of the so-called \textit{Fourth Paradigm} 
(i.e., data-intensive scientific discovery, \cite{fourthparadigm}).

APERTIF is expected to start its observing campaign of the northern sky in 2017. The 
daily APERTIF data cube will have dimensions of $2048\times2048\times16384$ $\sim 68.7 \times 
10^9$ voxels and the expected number of H\,{\small I} source detections is $\sim$ 100 every day. 
WALLABY will have similar characteristics. The large volume of data creates new 
needs, in terms of tools and algorithms which must exploit new ideas and solutions for storage, 
data reduction, visualization, and analysis to obtain scientific results. 

Visual analytics, \ref{ana}, the combination of automated data 
processing with human reasoning, creativity and intuition, supported by interactive 
visualization, is one of the prime methodologies that allow putting the human in the 
investigation loop. In this paper, we defined the visualization prerequisites and future 
perspective for applying this paradigm to H\,{\small I} observations focusing on the introduction of 3-D 
visualization in the process of source finding and analysis. In fact, the current astronomy 
visualization software has very limited 3-D capabilities for grid data (section \ref{astro}); 
while general purpose visualization software (section \ref{prere}, \ref{review_sof}) is not 
aimed at the analysis of H\,{\small I} data.

In this paper we showed:

\begin{enumerate}[i)]
\item more than 99$\%$ of the voxels in the H\,{\small I} datasets that APERTIF will deliver is 
dominated by noise and the sources are hidden in it (see Fig.\ref{fig0}). The current source 
finder software can extract them with high reliability and completeness \citep{Whiting, sofia}. 
The typical volume of individual sources will be ${50}^3 = 1.25 \times 10^5$  voxels (up 
to $512^3 \sim 1.3 \times 10^8$ in the case of occasional large galaxies), reducing the 
storage, I/O bandwidth and computational demands for visualization to a level accessible on 
desktops and laptops. The predicted weekly data rate, on the other hand, is high ($\sim 10^3$ 
sources). Fortunately, only a subset of these (2-3 sources per day) will be highly resolved (more 
than 10 resolution elements) or show complex features such as tails and extra-planar-gas. A 
powerful interactive visualization tool will be needed for fast inspection and analysis of 
these objects.

\item the analysis of the sources, for example producing moment maps and rotation curves, will 
also be done in an automated way. In particularly complex cases, human interaction will be 
necessary to drive the automated algorithm in the data volume and provide immediate feedback on 
the quality of the results (see section \ref{analysis}). Visualization tools with supervised 
semi-automated analysis algorithms will be needed. In fact, it becomes necessary to produce 
refined data with minimal time but maintaining the same level of quality. For example, the 
derivation of the rotation curve of a galaxy passes through the creation of the so-called 
tilted-ring model which, then, is compared to the data. This process has been converted to an 
automatic algorithm. However, significant kinematic features different from the Keplerian 
rotation (e.g., tidal tails, see Fig.\ref{fig01}) will be present in part of the data. The 
current algorithms can not automatically flag these features for the analysis. Therefore, human 
intervention is necessary to separate the regularly rotation disk and different kinematic 
features, and to feed the fitting algorithm with the selection, so that the user can judge the 
results quantitatively.

\item in section \ref{3-D}, we showed that 3-D visualization can enable an immediate overview of 
the kinematics of a galaxy, leading to improved understanding of the coherence in the data. 
Moreover, a high level of interactivity in all visualization aspects (rendering, smoothing, 
retrieving quantitative information, and comparative features) will be the key for enabling a 
fast inspection of the data. On the other hand, volume rendering has its 
limitations due to current 2-D input and output hardware. Some examples of these 
limitations are projection issues and the impossibility to move the cursor pixel by pixel. 
Adding 1-D/2-D views linked to the 3-D representation resolves these limitations. The 
combination with high-dimensional visualization techniques, which can help in finding outliers 
and patterns in the oceans of data, is also necessary.

\item in section \ref{prere} we identified the requirements for the visualization and analysis 
of H\,{\small I} in galaxies: interactive visualization with quantitative and comparative capabilities 
with 3-D selection techniques and supervised semi-automated analysis. Moreover, the source code 
must have the following characteristics for enabling collaborative work: open, modular, well 
documented, and well maintained. After a study of the state of-the-art of the open-source and 
actively maintained visualization packages with rendering of grid data capabilities (see 
section \ref{review_sof}), we adopted $\tt{3DSlicer}$ as a platform for developing a fully 
interactive desktop H\,{\small I} data visualization tool with quantitative and comparative features 
(section \ref{sli}). These techniques can also be used for other astronomical datasets such as 
3-D datasets provided by recent Integral Field Unit (IFU) observations \citep{califa, Wouter, 
Richard}. In that case, collaborative work is necessary to identify the key features needed to 
provide quantitative visualization.

\end{enumerate}

In conclusion, the success of a visualization tool depends heavily on the number of people using 
it over its life time. The life time of a software package depends on several factors such as 
usability, maintainability, and whether it has been developed with good insight in the subtle 
aspects of the data and its interpretation. $\tt{KARMA}$ is a perfect example of a successful 
package, developed in the mid 90's but still widely used by radio astronomers to date. Our aim 
is to achieve an analogous result exploiting the current hardware and algorithmic paradigms, 
focusing on the linking between 2-D and 3-D visualization, quantitative/comparative features and 
high-dimensional visualization.

\section{Acknowledgments}
Two of the authors, D. Punzo and J.M van der Hulst, acknowledge support from the European 
Research Council under the European Union's Seventh Framework Programme (FP/2007-2013) / ERC 
Grant Agreement nr. 291531. We thank R. Sancisi and F. Fraternali for very useful 
feedback. We also thank E. di Teodoro for providing $\tt{^{\rm 3D}\,Barolo}$ to us. 
Finally, we thank the reviewers for their constructive comments, which helped us to improve the paper 
substantially. Figures and videos in this paper were generated by using $\tt{3DSlicer}$ 
(\url{http://www.slicer.org/}).

%only for online version

\section{References}
\bibliographystyle{elsart-num-names}
\bibliography{visu.bib}

\newpage
\section{Additional on-line material}

In this section we provide videos of the volume rendering of part of the data presented in this paper \footnote{\url{https://www.youtube.com/watch?v=sS_5LrOS5bo}} \footnote{\url{https://www.youtube.com/watch?v=yLjW9nbdO8g}}.

\begin{figure}[h!]
\centering

\includegraphics[width=0.464\textwidth]{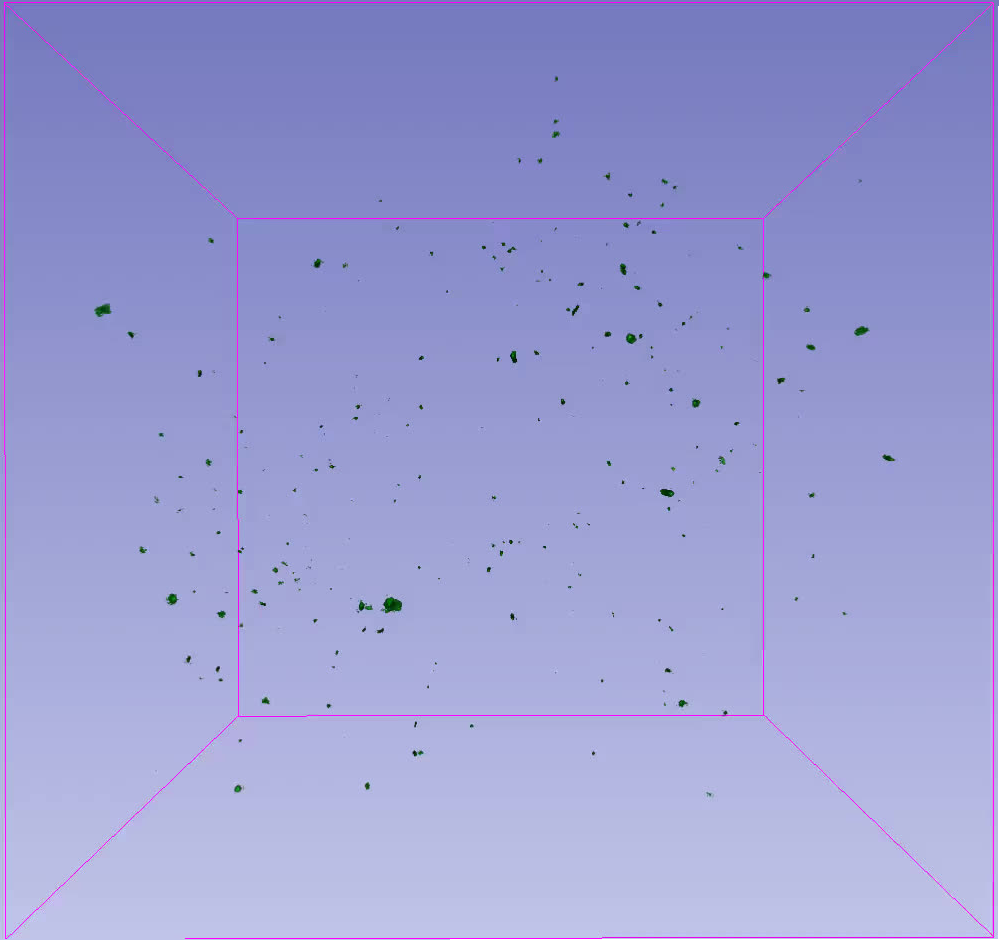}
%link for video_Fig1.mp4   

\includegraphics[width=0.464\textwidth]{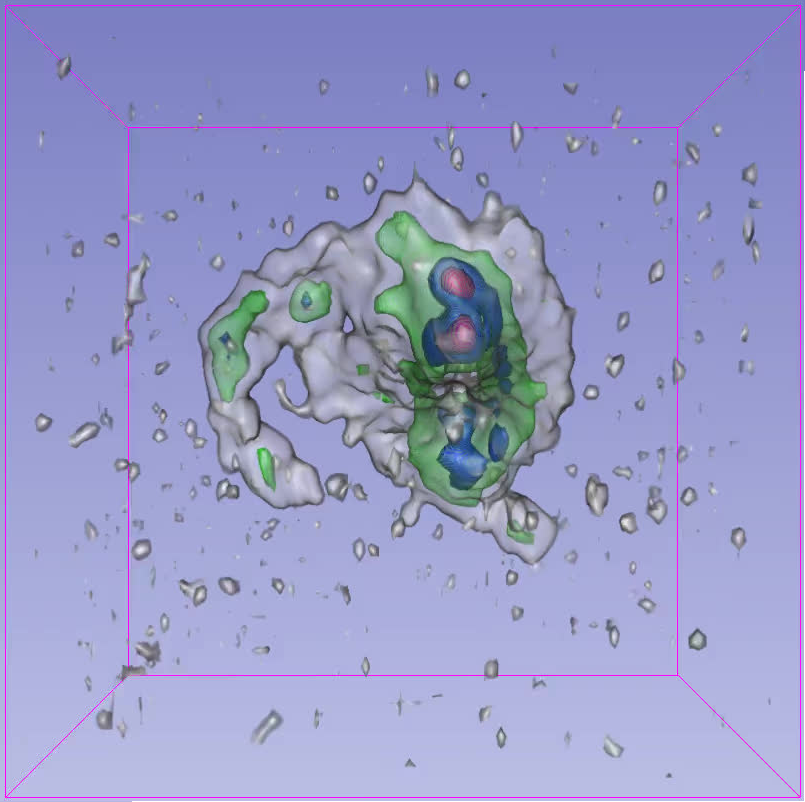}
%link for video_Fig2.mp4

\caption{Volume rendering of the data of the top-right panel of Fig.\ref{fig0} 
(top panel)  and of Fig.\ref{fig01} (bottom panel)  are shown. }
\label{figonline}
\end{figure}

\end{document}